%% file: main.tex
\documentclass[lettersize,journal]{IEEEtran}
\usepackage{amsmath,amsfonts}
\usepackage{algorithm}
\usepackage{array}
\usepackage{textcomp}
\usepackage{stfloats}
\usepackage{url}
\usepackage{verbatim}
\usepackage{graphicx}
\usepackage{cite}
\usepackage{orcidlink}
\usepackage{graphicx}
\usepackage{subcaption}

\newcommand{\name}{\textsc{DyPe}}

\usepackage[font=small,skip=2pt]{caption}
\usepackage{multirow} %multirow for format of table
\usepackage{colortbl}
\usepackage{xcolor}
\definecolor{yellow}{rgb}{0.95,0.95,0.53} %1,1,0.83
\newcolumntype{a}{>{\columncolor{yellow}}c}
\usepackage{algpseudocode}
\usepackage{hhline}
\usepackage{listings}
\usepackage{multirow}
\usepackage[symbol]{footmisc}

\newcommand{\dan}[1]{\textcolor{black}{#1}}%
\newcommand{\zhen}[1]{\textcolor{black}{#1}}%

\newcommand{\commentgreen}[1]{\textcolor{green!50!black}{\texttt{#1}}}

\usepackage{todonotes}

\begin{document}

\title{Data-aware Dynamic \zhen{Execution} of Irregular Workloads on Heterogeneous Systems}


\author{Zhenyu Bai~\orcidlink{0000-0003-1143-0762}, Dan Wu~\orcidlink{0009-0003-5260-0980}, Pranav Dangi~\orcidlink{0009-0004-1339-6048},
%~\IEEEmembership{Student Member,~IEEE}, 
Dhananjaya Wijerathne~\orcidlink{0000-0003-3181-2514},
Venkata Pavan Kumar Miriyala~\orcidlink{0000-0002-4984-4517}, and Tulika Mitra~\orcidlink{0000-0003-4136-4188} 
%~\IEEEmembership{Member,~IEEE}
\thanks{Zhenyu Bai, Dan Wu, Pranav Dangi, and Tulika Mitra are with the Department of Computer Science, National University of Singapore, Singapore. Emails: \{baizy, danwu20, dangi, tulika\}@comp.nus.edu.sg.}
\thanks{Dhananjaya Wijerathne and Venkata Pavan Kumar Miriyala are with AMD, Singapore. Emails: \{dmd.wijerathne, venkatapavankumar.miriyala\}@amd.com.}
}

 % \author{\phantom{Zhenyu Bai\footnotemark[2]\footnotemark[4], Dan Wu\footnotemark[2]\footnotemark[4], Pranav Dangi\footnotemark[2]\footnotemark[4], Dhananjaya Wijerathne\footnotemark[4]\footnotemark[1],} \\
 % \phantom{Dhananjaya Wijerathne\footnotemark[3]\footnotemark[1], 
 % Venkata Pavan Kumar Miriyala\footnotemark[3], Tulika Mitra \footnotemark[4] }}

\markboth{}%IEEE Transactions on Computers
{Bai \MakeLowercase{\textit{et al.}}: Data-aware Dynamic Execution of Irregular
Workloads on Heterogeneous Systems}


\maketitle

\begin{abstract}
\dan{
% Modern high-performance computing (HPC) systems handle both regular (e.g., dense tensor operations) and irregular computations (e.g., sparse tensor operations). While GPUs excel at performing regular, compute-intensive tasks, FPGAs and ASICs offer superior efficiency for irregular workloads. To leverage the strengths of both, heterogeneous servers combining GPUs with specialized accelerators have emerged as a promising solution. However, current approaches to scheduling workloads on such heterogeneous systems often rely on manual partitioning, offloading only the sparse computations to accelerators. 
Current approaches to scheduling workloads on heterogeneous systems with specialized accelerators often rely on manual partitioning, offloading tasks with specific compute patterns to accelerators. 
This method requires extensive experimentation and human effort to identify the tasks suitable for the accelerator.
% , and it lacks general applicability to workloads with a mix of regular and irregular computations.
}
\zhen{To solve this problem, we introduce {\name}, a scheduling framework tailored for heterogeneous systems with specialized accelerators. Our method automatically partitions, deploys, and reschedules execution when necessary by dynamically analyzing the characteristics of the input data and leveraging the interoperator parallelism among heterogeneous devices.}

{\name} navigates a multi-objective, multi-constraint design space that considers both system constraints and application requirements, which allows it to discover Pareto-optimal mapping configurations, \zhen{improving the system's overall performance and }effectively managing energy-performance trade-offs. To demonstrate the benefits of our approach \zhen{on real hardware}, we build a heterogeneous system of GPUs and FPGAs \zhen{with peer-to-peer data transfers}. The experiments show that conventional static scheduling is optimal for 13 out of 86 cases for different workloads and system settings while {\name} is adaptable and able to find the optimal schedule in 77 out of 86 cases, with an average of only 3.95\% performance or energy efficiency loss in the sub-optimal cases. Performance evaluation of {\name} shows an average of 1.53x throughput and 1.09x energy efficiency improvement over the static schedule baseline and 1.44x throughput and 1.66x energy efficiency over the GPU-only baseline.

\end{abstract}


\section{Introduction}
\label{sec:intro}

Sparse tensors play an important role in various high-performance computing (HPC) applications, including graph analytics, AI models, and scientific computation~\cite{han2015deep, gin, grossman2016survey}.
The source of sparsity can be graph structures, network pruning, sparse contexts in attention mechanisms, etc.~\cite{sze2017efficient, dave2021hardware, bai2024swat, butterfly}. The sparsity of data exhibits variability across different workloads or datasets, ranging over many orders of magnitude~\cite{ heterogeneous-chip}.

% Modern workloads, especially those machine learning, often consist of a diverse mix of dense and sparse computations.
Although GPUs are highly efficient for accelerating dense and regular workloads \zhen{due to the presence of dedicated matrix multiplication units} \dan{and SIMD execution paradigm}, they may encounter memory bottlenecks and lower compute utilization when dealing with sparse and irregular computations~\cite{sparse-computation-1,sparse-computation-2}. 
Conversely, accelerators such as FPGAs (Field-Programmable Gate Arrays) and ASICs (Application-Specific Integrated Circuits) have demonstrated superior efficiency in handling sparse and irregular memory-intensive workloads~\cite{firestone2018azure, bai2024swat}. 

%Recognizing this advantage, major cloud service providers like Microsoft Azure~\cite{ms-catapult} and Amazon AWS~\cite{amazon-aws} offer various platforms equipped with FPGA support. 
As a concrete example, our experiment shows that three AMD ALVEO\textsuperscript{\texttrademark} U280 FPGAs can deliver performance comparable to that of one
AMD Instinct\textsuperscript{\texttrademark} MI210 GPU when running the Sparse Matrix-Matrix Multiplication (SpMM) kernel with high sparsity level, achieving 1.6x greater energy efficiency than the GPU. Notably, this energy efficiency advantage of FPGAs over GPUs increases as the sparsity of the data increases, showcasing their suitability for handling highly sparse data common in certain neural network computations and large-scale data processing tasks. This underscores the potential advantages of heterogeneous systems, which capitalize on the complementary strengths of each type of device.

\begin{figure}
    \centering
    \includegraphics[width=\linewidth]{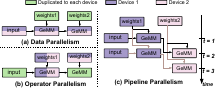}
    \caption{Different parallelism patterns.}
    \label{fig:parallelisms}
    \vspace{-10pt}
\end{figure}

\dan{To fully leverage resources in such a multi-machine system, different parallelism strategies can be employed.} \zhen{The most common approach is \textit{data parallelism}\cite{dean2012large}, illustrated in Figure~\ref{fig:parallelisms}a. In data parallelism, the entire model (including all weights) is duplicated across multiple devices, and each device executes the model independently on different parts of the input data or on different input batches~\cite{virtualflow, elasticflow, optimus}. \textit{Operator parallelism}, depicted in Figure~\ref{fig:parallelisms}b, allows different devices to collaboratively execute a single operator in parallel, effectively splitting the computation of that operator across devices. However, these parallelism paradigms are not efficient in heterogeneous systems where different devices are better suited for executing different operations. In such scenarios, \textit{inter-operator parallelism}~\cite{zheng2022alpa}, or \textit{pipeline parallelism} (Figure~\ref{fig:parallelisms}c, becomes essential. Pipeline parallelism assigns different operators (or layers) of the model to different devices, enabling each device to process its assigned operation and pass the intermediate results to the next device in the pipeline.}

\dan{Existing works \cite{jiang2021fleetrec, li2022hyperscale, zhang2022low} } \zhen{ focus on scheduling a given workload with a known performance profile} \dan{to the heterogeneous systems}. \zhen{Therefore, practical implementations} predominantly employ \textbf{static} resource allocation, where a fixed number of devices are dedicated to specific parts of the workload \zhen{that are ad-hoc to specific workloads and system configuration} \cite{li2022hyperscale, zhang2022low}. FleetRec~\cite{jiang2021fleetrec} advances further by enabling dynamic adjustment in the number of allocated devices while maintaining a static selection of device types. Such approaches, however, rely on two assumptions that may not always be valid: (1) the characteristics of the workload are fully known at design time and remain unchanged, and (2) \dan{the performance of a single device type is consistently superior, diminishing the need to offload tasks to alternative device types.}
These assumptions can result in suboptimal resource usage, particularly as the performance characteristics associated with different input data—such as input dimensions and arithmetic intensities—or system characteristics like internal data transfer bandwidth evolve dynamically. 
Consequently, no single schedule or resource allocation guarantees superior performance or energy efficiency universally. This underscores the need for more adaptive scheduling approaches that can dynamically adjust resource allocations and schedules in response to real-time changes in workload characteristics. 

\begin{figure}[h]
    \centering
    \includegraphics[width=\linewidth]{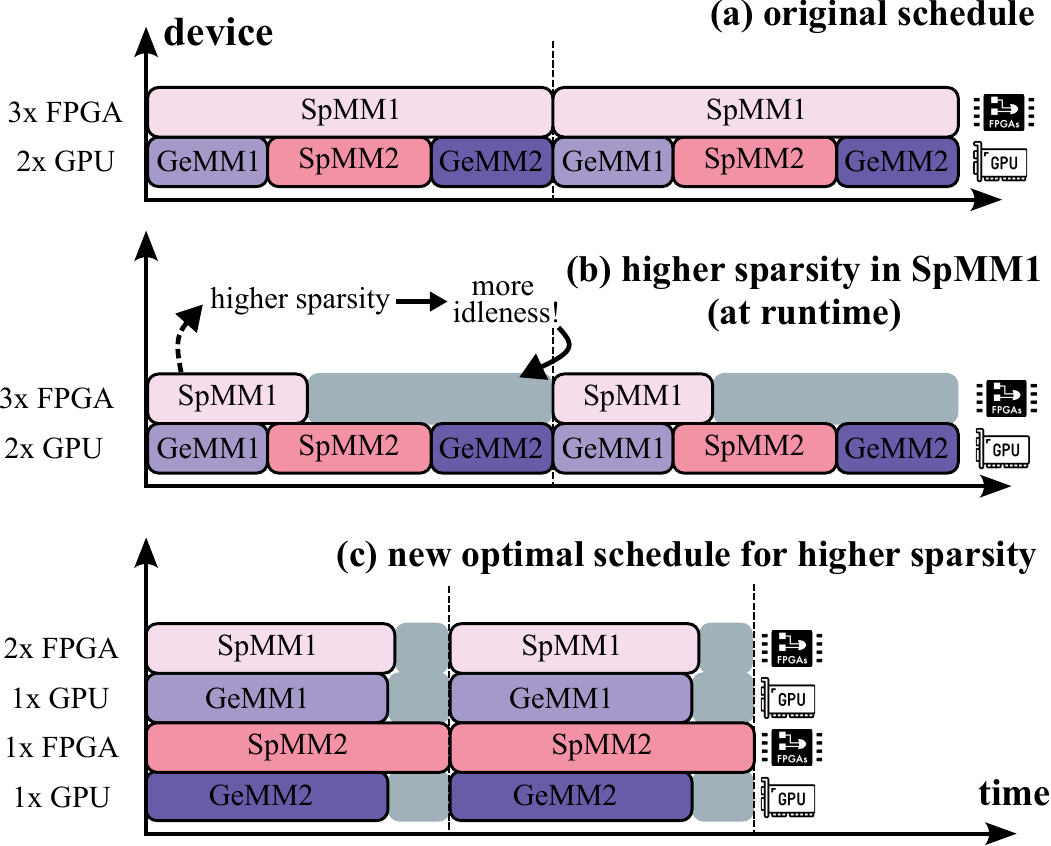}
    \caption{Example pipelined schedules for GCN inference. (a) 2-stages example, (b) same schedule with higher sparsity in SpMM1, and (c) the new optimal schedule considering the sparsity change.}
    \label{fig:pipeline-examples}
    \vspace{-10pt}
\end{figure}

\zhen{The combination of variability in data characteristics, heterogenity, and the necessity of pipeline parallelism creates even more chanllenges for efficient scheduling. As shown in the motivating example,}
Figure~\ref{fig:pipeline-examples}a presents a static schedule for a 2-layer Graph Convolution Network~(GCN) inference process on a heterogeneous system with two GPUs and three FPGAs. Each layer involves a Sparse Matrix-Matrix Multiplication (SpMM) kernel capturing the graph structure, followed by dense kernels, General Matrix Multiply (GeMM), for vertices' feature extraction. For clarity, we annotate each kernel in the model with its type and the layer in which it resides, namely SpMM1, GeMM1, SpMM2, and GeMM2.
Figure~\ref{fig:pipeline-examples}b shows the execution profile of the system with higher sparsity in the SpMM kernel, which reduces its execution time. \zhen{Considering that such changes in data characteristics are often statically unpredictable, the static schedule now leads to inefficient resource usage due to the imbalanced execution times of pipeline stages.} Figure~\ref{fig:pipeline-examples}c presents a re-optimized schedule that adapts to the increased sparsity by altering device allocations. This adaptation enhances overall system performance by ensuring that all resources are utilized efficiently, minimizing idle time, and aligning execution times more closely across different stages.

Finally, achieving the highest throughput is not always the only objective. For example, achieving a specific Quality of Service (QoS) target or optimizing a combination of multiple objectives, such as minimizing energy consumption after achieving a certain throughput, can also be essential considerations~\cite{japan-supercomputer}. Heterogeneity enables trade-offs between performance and energy efficiency. When several types of devices can be used for one kernel, they show different performance-energy tradeoffs. For example, FPGA shows better energy efficiency but worse throughput for the SpMM kernel in some of our experiments.  The different performance-energy profiles, together with the device allocation strategy, can help in achieving the right trade-off.

To address the above issues, we introduce a novel dynamic scheduling framework for heterogeneous systems, {\name}. This lightweight, input-aware framework operates within a multi-objective, multi-constraint design space, aiming to generate Pareto-optimal schedules that effectively manage the dynamic nature of applications and systems. We address three key challenges in {\name} to achieve our objective: (1) dynamically creating optimal schedules in a heterogeneous environment with respect to the characteristics of the system and data input to meet the design objectives; (2) accurate and efficient estimation of the compute kernels performance across diverse devices to make informed workload offloading decisions on  heterogeneous devices, and (3) a proof-of-concept cluster equipped with direct FPGA-GPU communication to validate our approach.

We demonstrate the benefits of {\name} with two substantial case studies that include both sparse and dense computations: Graph Neural Network (GNN)~\cite{GNN} and Sliding-window-based Transformers~\cite{attention-is-all-you-need,longformer}. {\name} finds the optimal schedule for 77 cases out of the 84 different system settings and workload combinations. The average throughput or energy efficiency loss in the 7 sub-optimal cases is limited to only 5.94\% and 2.46\%, respectively.
Performance evaluation of {\name} shows an average of 1.53x throughput and 1.09x energy efficiency improvement over the manually tuned static schedule baseline, and 1.44x throughput and 1.66x energy efficiency over the GPU-only baseline.

\section{{\name} Framework}
\begin{figure}[h]
    \centering
    \includegraphics[width=\linewidth]{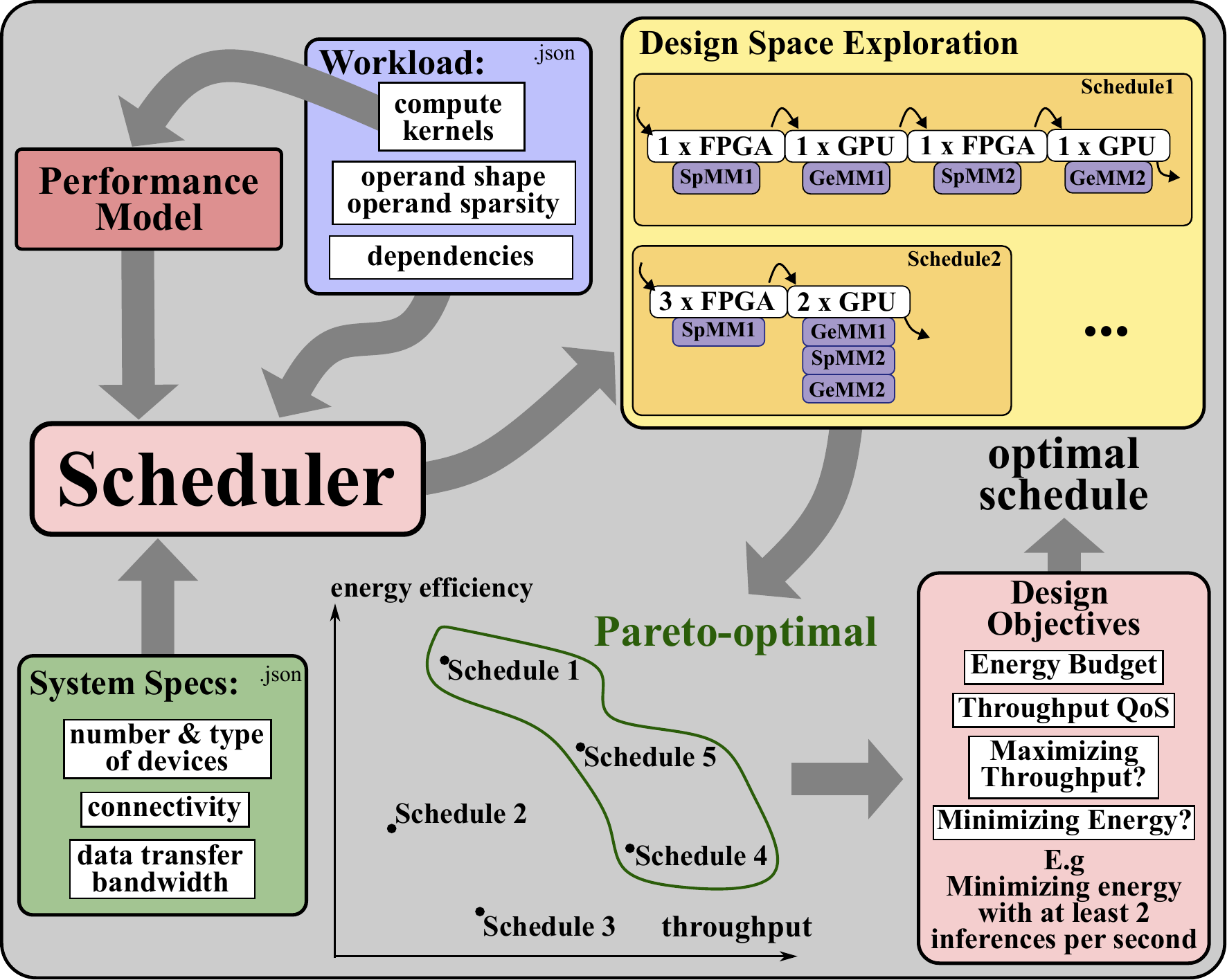}
    \caption{The {\name} framework.}
    \label{fig:dype-overview}
    \vspace{-10pt}
\end{figure}

Figure~\ref{fig:dype-overview} illustrates the {\name} framework, where the scheduler forms its core. Inputs to the scheduler include: 1) target workload description, 2) system specifications, 3) kernel performance model, and 4) design objectives.

\noindent\textbf{Target Workload} comprises compute kernels characterized by input dimensions, sparsity, and dependencies. These properties guide device selection (e.g., lower sparsity in kernels may favor GPUs over FPGAs). The workload characteristics impact kernel execution times and pipeline balance, where imbalance can lead to increased stalling, reducing system efficiency.

\noindent\textbf{System Specifications} encompass device count, types, interconnections, and data transfer capabilities. The scheduler factors in these when mapping devices to compute kernels, accounting for data transfer costs among devices.

\noindent\textbf{Kernel Performance Model} aids in estimating execution times and requires both precision and efficiency. Our approach employs a two-step model: first, generating synthetic input profiles and benchmarking kernel performance; second, training a linear regression model with this data. While linear models provide a foundation, the framework can incorporate more detailed models for complex kernels as discussed in Section~\ref{sec:perf-model}.

\noindent\textbf{Design Objectives} set user-defined goals, such as maximizing throughput or minimizing energy while maintaining a minimum required throughput. This flexibility supports various energy-performance trade-offs, aligning with specific operational needs.

The Scheduler explores design configurations, seeking Pareto-optimal schedules to meet workload and system constraints. The scheduler can dynamically adapt to new scenarios, as in GNN applications like traffic forecasting, social network recommendations, and graph analytics. For example, in traffic forecasting, {\name} can employ a high refresh rate for traffic flow predictions during peak hours for greater accuracy and reduce energy consumption during off-peak times.

\newcommand{\Ops}{\mathcal{OP} }
\newcommand{\Pipes}{\mathcal{PIPE} }
\newcommand{\concat}{^\frown}
\newcommand{\Nat}{\mathbb{N}}
\newcommand{\Devices}{\mathcal{DEV}}

\input{scheduling_new}

\section{System Design}

\subsection{Hardware Setup}
\label{sec:setup}
\begin{figure}[t]
    \centering
    \begin{subfigure}{0.49\columnwidth}
        \centering
        \includegraphics[width=\columnwidth]{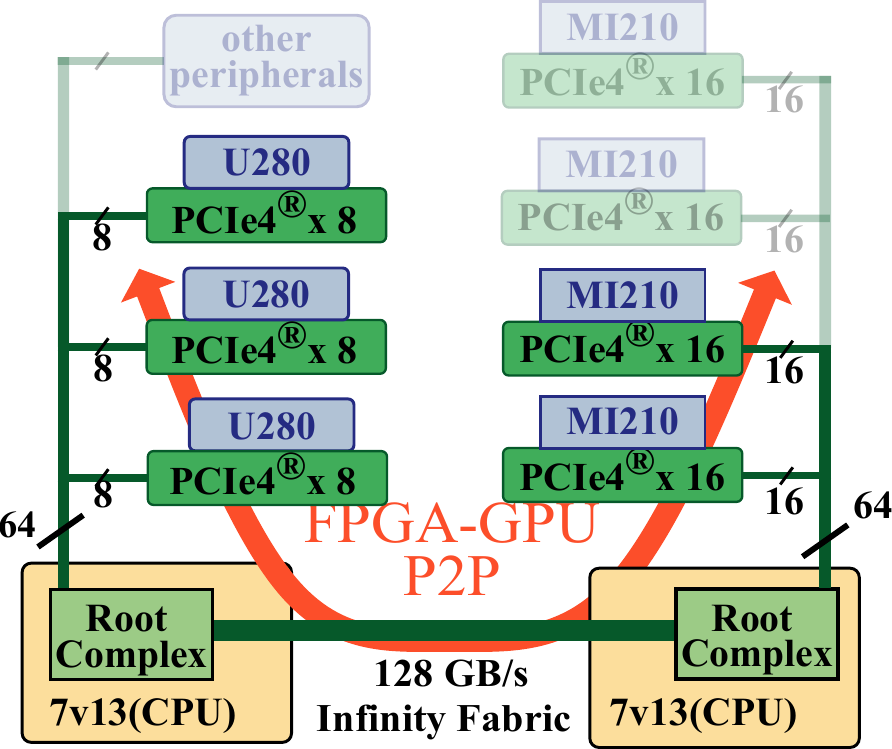}
        \caption{System topology.}
        \label{fig:system-topology}       
    \end{subfigure}
    \begin{subfigure}{0.49\columnwidth}
        \centering
        \includegraphics[width=\columnwidth]{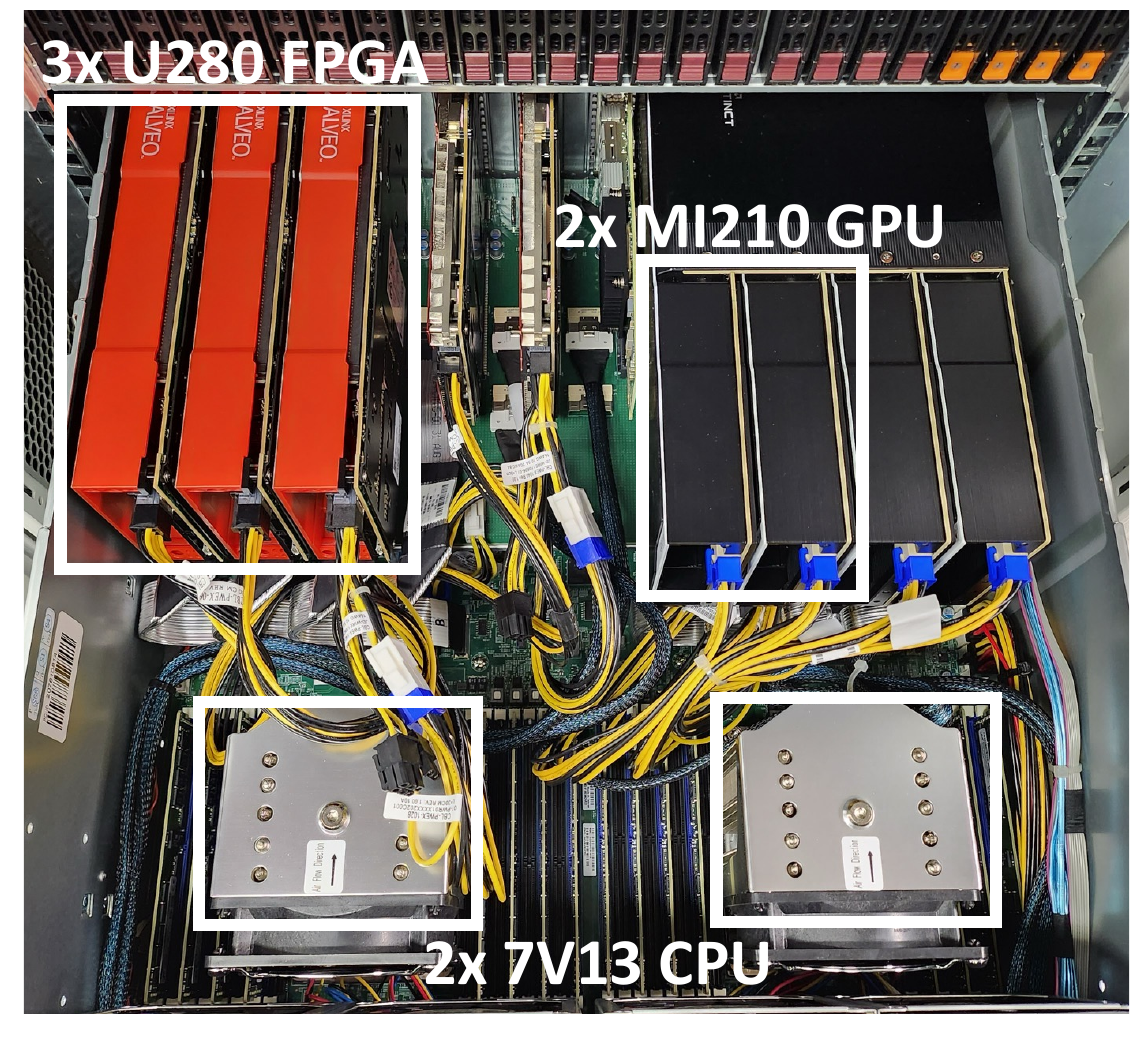}
        \caption{Photo of the testbed.}
        \label{fig:cluster-photo}   
    \end{subfigure}
    \caption{System hardware}
    \label{fig:system}     
    \vspace{-8pt}
\end{figure}

To ensure realistic validation, we built a heterogeneous cluster containing AMD EPYC\textsuperscript{\texttrademark} CPUs, AMD Instinct\textsuperscript{\texttrademark} MI210 GPUs, and AMD ALVEO\textsuperscript{\texttrademark} U280 FPGAs for system-level performance evaluation.
% We built a heterogeneous cluster using AMD EPYC\textsuperscript{\texttrademark} CPUs, AMD Instinct\textsuperscript{\texttrademark} MI210 GPUs, and Xilinx U280 FPGAs. 
Figure~\ref{fig:system} shows the cluster installation and the network topology among the components. Two AMD EPYC\textsuperscript{\texttrademark} 7v13 64-core CPUs are at the center of the system, and each of them integrates a root complex managing 128 PCIe\textsuperscript{\textregistered} lanes. Half of the 128 PCIe\textsuperscript{\textregistered} lanes are used for CPU-CPU connection, enabling a bandwidth of 128 GB/s. The remaining 64 lanes connect peripherals, including the accelerators. The first CPU manages four MI210 GPUs, each using 16 PCIe\textsuperscript{\textregistered}4.0 lanes (31.52 GB/s physical bandwidth). Note that we used only two GPUs for this study.
%to hardware sharing policies \intodo{what does this mean? hardware sharing policies?}, only two GPUs are available for our experiments. 
The second CPU hosts three U280 FPGAs, each utilizing 8 PCIe\textsuperscript{\textregistered}4.0 lanes (half of the lanes compared to GPUs, 15.76 GB/s physical bandwidth).

\subsection{FPGA-GPU Peer-to-Peer Data Transfer}
\label{sec:p2p}
Data transfer among accelerators is critical in our heterogeneous system. Traditionally, transferring data between accelerators involves routing through the CPU main memory, resulting in unnecessary data transfer overhead. To avoid this, we enabled the FPGA-GPU Peer-to-Peer (P2P) data transfer. We achieve this by mapping the FPGA memory onto the PCIe\textsuperscript{\textregistered} bus so the GPUs can be the masters of the data transfers and can read or write directly to the FPGA's memory~\cite{fpga-gpu-p2p}. The transfers are physically managed by the root complexes of the CPUs and the memory controllers of GPUs and FPGAs, as shown in Figure~\ref{fig:system-topology}, which are configured using the Xilinx Runtime~(XRT)%~\cite{amdxrt} 
OpenCL extension by address mapping. 
% \todo[inline]{We can say a bit more about that}
We discovered that the read or write operations on the GPU side can be achieved using the legacy instructions for GPU-CPU data transfer (\texttt{hipMemcpy()}) as the destination or source pointer points to a correctly mapped address. This setup enables multiple GPUs to transfer data to several FPGAs concurrently through the PCIe\textsuperscript{\textregistered} bus. The overall bandwidth is determined by the combined bandwidths of the involved GPUs and FPGAs. 

\begin{figure}
    \centering
    \includegraphics[width=\linewidth]{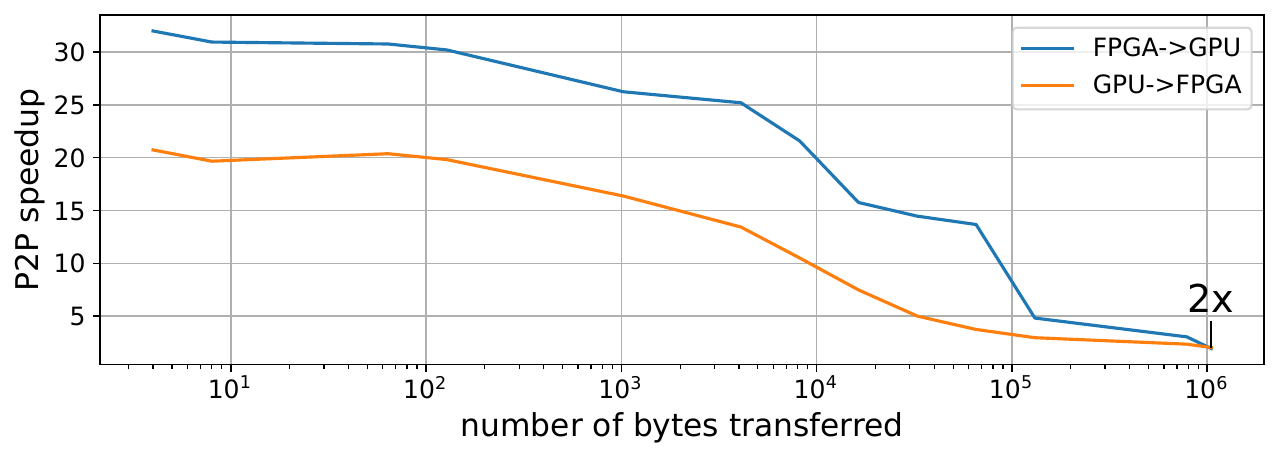}
    \caption{Data transfer speedup with P2P direct data transfer.}
    \label{fig:p2p-analysis}
    % \vspace{-10pt}
\end{figure}

To validate the benefits of P2P data transfer, we conducted a comparative analysis between traditional data transfers involving the CPU and direct P2P transfers. The results, depicted in Figure~\ref{fig:p2p-analysis}, show that CPU involvement introduces considerable overhead, especially with smaller data amounts. As the total transfer size increases, the speedup from using direct P2P transfers converges to approximately 2x for transfers of 1MB. This substantial reduction in overhead underscores the effectiveness of the P2P approach.

\section{Workloads Case Studies}

% The {\name} framework leverages configuration files written in JSON to describe workloads, offering flexibility in adapting to various computational tasks and workload characteristics. This design facilitates the straightforward integration of new workloads and ensures that {\name} can be easily adapted to support emerging applications that may benefit from a heterogeneous computing setup.

% The structure of the {\name} framework is designed to handle streaming multiple inferences (or their equivalents) over a given workload. This means that, while being monitored, whenever there is a \textit{significant} change to any aspect of the application or system or a new workload needs to be executed, {\name} can dynamically produce a new schedule. 
% %The term "significant" is emphasized because {\name} intelligently relies on estimation models, allowing it to tolerate minor changes or iterative updates to input characteristics without affecting the schedule. 
% This capability makes it even more suitable for downstream tasks with iterative updates where schedule refreshes are required only after many changes have accumulated. i.e., a schedule refresh can be triggered after every specified number ($n$) of inferences or at regular time intervals ($t$ minutes).

For the scope of this work, as a case study, we evaluate {\name} using two representative target workloads: (a) inference for GNNs and (b) inference for transformers employing sliding-window attention. These examples are specifically chosen due to their prevalent use in current high-performance computing tasks, their distinctive mix of regular and irregular computations\zhen{, and their dynamic nature in the real-world applications}.

%For instance, in GNN applications, the {\name} framework is particularly well-suited for traffic forecasting, social network recommendations, and graph analytics tasks. In these scenarios, the sparse input matrices typically exhibit consistent high-level characteristics (e.g., dimension and sparsity), necessitating infrequent refreshes to the schedule. The various operational modes of DyPe are particularly beneficial in applications such as traffic forecasting, which is essential for online maps. By utilizing DyPe in performance-optimization mode, online maps can use a higher refresh rate for predicted traffic flows, leading to more precise estimated arrival times and quicker responses during peak hours. Conversely, during non-peak times, a balanced mode of DyPe can be employed to conserve energy with an acceptable refresh rate.

%It is important to highlight that while our focus is on these two applications, the {\name} framework is designed with a flexible architecture that accommodates a broad range of other computational tasks. This flexibility is achieved through a system that interprets workload descriptions provided in configuration files. This design not only facilitates the straightforward integration of additional workloads but also ensures that {\name} can be readily adapted to support emerging applications that may benefit from its dynamic scheduling capabilities and efficient resource management.

\subsection{GNN workloads}
GNNs have gained immense popularity due to their effectiveness in modeling relational data, with applications in diverse areas, including social network analysis and recommendation systems. 
Among these, graph convolutional networks (GCN)~\cite{gcn} are prominently used for node classification and graph segmentation, while graph isomorphism networks (GIN)~\cite{gin} are employed for tasks requiring intricate graph representations, such as characterizing complex interactions within social networks and traffic forecasting.
GNN workloads are composed of both neural network operations and graph-based irregular sparse operations for feature extraction, which makes them perfect candidates for investigating the potential of heterogeneous systems.

One GCN layer can be represented as Equation \ref{eq:gcn}:
\begin{equation}\label{eq:gcn}
    \mathbf{X}' = \mathbf{\hat{A}}\mathbf{X}\Theta
\end{equation}
where $\mathbf{\hat{A}} = \mathbf{D}^{-\frac{1}{2}}(\mathbf{I}+\mathbf{A})\mathbf{D}^{-\frac{1}{2}}$ denotes the adjacency matrix with inserted self-loops normalized by vertex degree,
% $D_{ii}=\sum_{j}\hat{A}_{ij}$ is the degree matrix, 
%$X$ is the matrix of feature or hidden states, 
and $\Theta$ is the trained weight for linear transformation.
% and ReLU is the activation function.
% GCNs are successful in graph learning because they can capture graph structural information and task-depend information through the adjacent matrix and trainable weights separately.
Since the adjacent matrix is highly sparse while feature and trained weights are usually sparse, GCN can be partitioned into two parts: a SpMM kernel for $\mathbf{Y}=\mathbf{\hat{A}}\mathbf{X}$, and a GEMM for $\mathbf{X'}=\mathbf{Y}\Theta$.

One GIN layer can be represented as Equation \ref{eq:gin}:
\begin{equation}\label{eq:gin}
    \mathbf{X}' = MLP(\mathbf{A'} \mathbf{X})
\end{equation}
where $\mathbf{A'} = \mathbf{A}+(1+\epsilon)\mathbf{I}$ denotes the adjacency matrix with inserted self-loops times a learned parameter $\epsilon$,
% $D_{ii}=\sum_{j}\hat{A}_{ij}$ is the degree matrix, 
%$X$ is the matrix of feature or hidden states, 
and $MLP$ is a trained multilayer perceptron (MLP).
A $n$-layer MLP is equivalent to $n$ consecutive GEMM kernels.
Therefore, one GIN layer can be represented as a SpMM kernel for $\mathbf{Y}=\mathbf{A'}\mathbf{X}$ followed by multiple GEMM kernels for the MLP.

We use two GNN models, GCN~\cite{gcn} and GIN~\cite{gin}, as benchmark workloads.
Both models have two layers with a hidden state length of 128. Table~\ref{tab:dataset} shows the datasets used in the experiments.  The {synthetic 1-4} datasets are synthetic datasets to improve the diversity of the workloads' characteristics. %The following ogbn- datasets are from the Open Graph Benchmark~\cite{open-graph-benchmark}.

\begin{table}[h] \small
\centering
\caption{Datasets for GNN workloads.}
\label{tab:dataset}
\resizebox{\linewidth}{!}{
\begin{tabular}{r|ccccc}
 \hline
   Dataset & \#Vertex & \#Edge & Sparsity& Feature Len.\\
 \hhline{=|====}
% PubMed (PM) &\cite{planetoid} & 19,717 & 88,648 & 500 \\
% Github (GH) &\cite{rozemberczki2021multi} & 37,700 & 578,006 & 128  \\
synthetic 1 (S1)  & 230K & 120M &99.77315\% & {600} \\
synthetic 2 (S2) & 230K & 15M & 99.95274\%  & {600} \\
synthetic 3 (S3)  & 700K & 15M & 99.99693\% & 300 \\
synthetic 4 (S4) & 3.5M & 5M& 99.99995\% & 20 \\
\hline
ogbn-arxiv (OA)\cite{open-graph-benchmark} & 170K & 1.1M & 99.99593\%      & 128 \\
%ogbn-mag & 1.9M & 21M & 0.9999943893\%      & 128 \\
ogbn-products (OP)\cite{open-graph-benchmark}& 2.4M & 61M & 99.99793\%      & 100 \\
% amazon-products (AP)\cite{zeng2019graphsaint} & 1.6M & 264M &99.98928\% & 200 \\

%DgraphFIN (DF)\cite{huang2022dgraph}& 3.7M & 4.7M& 99.99997\% & 24 \\
 \hline
\end{tabular}
}
\vspace{-10pt}
\end{table}

\subsection{Transformers with sliding-window attention}
\label{sec:sliding-window}

Irregularity also presents challenges in transformer models, mainly in attention head. For example,~\cite{butterfly} demonstrates the use of FPGAs to accelerate Discrete-Fourier-Transform based attention~\cite{fft-attention}. SWAT~\cite{bai2024swat} uses FPGAs to accelerate the sliding-window attention mechanism~\cite{longformer, bigbird}. Both are approximations of the vanilla attention mechanism. However, transformer models also incorporate regular operations such as linear projections and Feed-Forward Neural Networks (FFNs)~\cite{attention-is-all-you-need}, which are not addressed or not efficiently handled.
%Our work builds upon these findings by demonstrating that a hybrid approach, which combines the capabilities of FPGAs and GPUs, can enhance the performance of transformer model inference.

%The computational complexity of vanilla attention~\cite{attention-is-all-you-need} in transformer models is quadratic with respect to the length of the input, which significantly impacts execution time as input length increases \cite{transformer-survey}. To address this scalability issue, sliding-window attention mechanisms, such as those proposed in \cite{longformer, bigbird}, limit the attention each input token can attend to, thereby reducing the theoretical complexity to linear with respect to input length.

The computation within one vanilla transformer layer involves several key operations:
\begin{gather}
        Q = I \times W_Q, K = I \times W_K, V = I \times W_V \label{eq:linear-projection}\\ 
        S = Q \times K, S' = Softmax(S), Z = S' \times V \label{eq:attention}\\ 
        O = FFN(Z) \label{eq:FFN}
\end{gather}
The sliding-window attention technique limits each input token of $Q$ to attend to statically fixed tokens of $K$, which reduces the theoretical complexity to linear. This is equivalent of re-writing Equation \ref{eq:attention} as:
\begin{equation}
    S = \text{MASK}(Q \times K), S' = Softmax(S), Z = S' \times V \label{eq:window-attention}
\end{equation}

Where the mask applied to $Q \times K$ introduces a sparsity, making the $S = \text{MASK}(Q \times K) $ a Sampled-Dense-Dense-Matrix-Multiplication, and the $ Z = S' \times V $ an SpMM. The sparsity introduced by the mask means that these computations can be efficiently handled by FPGAs~\cite{bai2024swat}.
While SWAT~\cite{bai2024swat} presents an end-to-end FPGA-based solution for efficiently handling sliding-window attention with improved performance and energy efficiency, it primarily focuses on the attention computation and does not address the regular computations Equations \ref{eq:linear-projection} and \ref{eq:FFN}, which are better suited for GPUs. In our study, we extend the work of SWAT by utilizing {\name} that enables a hybrid approach for supporting transformers with sliding-window attention, leveraging both FPGAs and GPUs to maximize energy efficiency and throughput.

In our evaluation, we consider the classical transformer model. Reflecting the constraints outlined in \cite{bai2024swat}, we adopt the BigBird setting with an attention dimensionality of 512 with 8 heads. We consider a model of 32 layers, aligning with the Mistral-7B model with the sliding-window mechanism~\cite{mistral-7B}. The window size ($w$) is set between 512 and 4096, and the input length ($seq\_len$) is set between 1024 to 16384\footnote{while $w\leq seq\_len$ is ensured for a valid combination.}. $w$ and $seq\_len$ are treated as dynamic input characteristics, necessitating the {\name} scheduler to dynamically determine the optimal system configuration.

\section{Kernel Performance models}
\label{sec:perf-model}
To accurately and quickly estimate the execution time of the various kernels in our target workloads, we follow a two-step process to set up the performance model. First, we generate synthetic inputs reflecting a wide array of possible input characteristics. The kernel performance is measured on the hardware, forming a training dataset for known performance. 
Second, we employ these data points to train a linear regression model that predicts kernel performance.

While a simple linear model provides a good starting point, it may not always capture the complex relationships inherent in certain types of operations. To enhance the model’s accuracy, we incorporate more complex characteristics that offer a more detailed description of the workload inputs. For instance, in SpMM kernels, beyond just utilizing shapes and sparsity, we also include arithmetic intensity as an additional predictor. Arithmetic intensity, which represents a non-linear combination of shapes and sparsity, provides a deeper insight into the computational demands of the operation. 

In cases where more specialized estimation is required—such as our FPGA implementations, where specific rough performance models are known from the architecture, we use the rough performance formula as \zhen{one} input parameters of the linear regression model.

We detail the performance models for the kernels in our example workloads as follows:

\noindent\textbf{SpMM on GPU} has its performance predicted with the sparse matrix shape ($M, K$), number of non-zero values $nnz$, and the dense matrix shape ($K, N$) as input. We further use GFLOP and arithmetic intensity $arm$ as inputs to reflect the computation and computation-memory ratio separately.  $GFLOP= (2 nnz N-M N)\cdot10^{-9}$,  $arm= \frac{GFLOP \cdot 10^9}{8(nnz+MN)}$, and both are non-linear functions of nnZ and dimensions of inputs, which creates a more complex function of the inputs beyond the linear regression model.
The estimation formula is 
\begin{equation}
    t = C_1 \cdot N + C_2 \cdot nnz + C_3 \cdot GFLOP + C_4 \cdot arm
\end{equation}

\noindent\textbf{SpMM on FPGA} has predictable performance thanks to the timing predictability of FPGAs. \cite{song2022sextans} provides an original performance model. Considering our modifications on the original implementation\footnote{Original Sextans implementation supports general SpMM kernel in the form of ~~ $C = \alpha A B + \beta C$, we removed the $+\beta C$ and $\alpha$. The freed resources are used to increase the number of functional units}, we add a scaling factor $C$ to the original formula: %And apply the linear regression to find the best $C$.
\begin{equation}
t = C \cdot \frac{(nnz+13M)N}{F \cdot N_{M} \cdot 10^3}\notag
\end{equation}
where the operational frequency ($F$) = 215~MHz and the total number of multiplication and accumulation units ($N_{M}$) = 640 are inherited from the original paper. 
\\\noindent\textbf{GEMM on GPU} performs regularly with a strong correlation on the number of MAC and the actual GFLOP of the workload. 
The shape and the dimensions of inputs also have an impact due to the relationship to the size of the matrix multiplication intrinsic. 
Therefore, we mix all these parameters into the estimation formula:
\begin{equation}
\begin{aligned}
t = &C_1\cdot K + C_2 \cdot N + C_3 \cdot MN + C_4 \cdot MK \\
   &+ C_5 \cdot KN + C_5 \cdot MKN + b
\end{aligned}
%\begin{aligned}
%t = &2.5 \times 10^{-4} K  - 9.7 \times 10^{-4} N + 3.7\times 10^{-9} MN  - 1.6\times 10^{-9} MK\\ 
%&-4.2\times 10^{-6} KN + 9.0\times10^{-11} MNK + 0.15 \notag
%\end{aligned}
\end{equation}

\noindent\textbf{Sliding-Window on FPGA} also has predictable performance thanks to the predictability of FPGAs and the optimized memory bandwidth usage in the design of SWAT~\cite{bai2024swat}. We inherit the design of SWAT\cite{bai2024swat}, adding a scaling factor to the original performance model:

\begin{equation}
    t = C \cdot (seq\_len \cdot t_{pipeline} + t_{init}) \cdot (w / 1024) / F
\end{equation}
Where $t_{pipeline}= 201$ and $t_{init}=904$ and $F=421MHz$ are the design parameters in \cite{bai2024swat}.
$seq\_len$ and $w$ are the transformer model's sequence length and window size.

Each model uses coefficients $C_{\ast}$ that are trained on generated data points, enhancing the predictive accuracy of these performance estimations.

\noindent\textbf{Sliding-Window on GPU.}
GPUs generally struggle with the specialized sparsity patterns typical of sliding-window attention, as their architecture is not ideally suited for handling sparse data structures efficiently. Research indicates that the state-of-the-art GPU implementations of sliding-window attention found in platforms like HuggingFace and XFormers only reduce memory usage, with only minimal decreased execution time~\cite{bai2024swat, longformer}. 
Considering that the primary focus of this paper is on computational efficiency and scheduling rather than on optimizing specific kernels, we choose to base our GPU performance model on a standard dense computation approach. 

\section{Evaluation}
\subsection{Experiment Setup}
\label{sec:exp-setup}

\noindent\textbf{Baselines.}
We use three types of baselines: 1) homogeneous system baselines; 2) addition of homogeneous system result; 3) other scheduling strategies on heterogeneous system.

The homogeneous system baselines are \textit{GPU-only} and \textit{FPGA-only}.  For the \textit{GPU-only} and \textit{FPGA-only} methods, we use only a single type of device in our heterogeneous system and remove the rest. The specifications of each type of device are listed in Table~\ref{tab:system}.  

While comparing the energy efficiency of {\name} to the \textit{GPU-only} and \textit{FPGA-only} baselines is informative, it should be noted that these baselines have fewer computing resources. To ensure a fair comparison, we also introduce the \textit{theoretical-additive} baseline. This baseline simply sums the throughput of the \textit{GPU-only} and \textit{FPGA-only} configurations and averages their energy efficiency. It represents a simplistic model of a heterogeneous system where computing resources are uniformly distributed across all kernels, without considering the specific advantages of each device type.

We use two methods as the static schedules for heterogeneous system baselines: \textit{static} and \textit{FleetRec}.
The manually-tuned \textit{static} baseline follows a common scheduling approaches, which statically assigns specific devices to specific parts of the workload without flexibility. 
% For GNN workloads, \textit{static} assigns two FPGAs to the SpMM of the first layer, one FPGA to the SpMM of the second layer, and two GPUs to handle all GeMM kernels. In sliding-window-based transformer models, \textit{static} allocates all FPGAs for attention computations and GPUs for the remaining tasks. The \textit{FleetRec} baseline represents a slightly more flexible approach compared to \textit{static}, allowing adjustments in the amount of resources employed for the kernels but maintaining fixed types of devices for specific kernels.
We implemented \textit{FleetRec} with {\name} by applying design constraints to limit the fixed types of devices on specific kernels, hence referred to as \textit{FleetRec}$^*$ in the subsequent experiments.

% \intodo{cite} where computation goes to the most appropriate device type, we use a static scheduling approach called \textit{static}. FPGAs have better energy efficiency on SpMM (Figure ~\ref{fig:effect_of_sparsity}) and the first SpMM in GNN is heavier than the second in most cases. Therefore, 

%  SpMM to FPGAs and GEMM to GPUs in our context. However, given that GPUs can process SpMM (albeit with lower energy efficiency), there are scenarios where offloading SpMM to GPUs is advantageous for minimizing data transfers or balancing pipeline stages. This is an aspect incorporated into {\name}'s dynamic scheduling approach. To illustrate {\name}'s dynamic scheduling capabilities, we analyze a static schedule method, termed \textit{static}. This method dedicates 2 FPGAs for SpMM in the first GNN layer and 1 FPGA for the second layer's SpMM, based on the greater size of the first layer's SpMM in most datasets.
% The \textit{static} approach also implements system pipelining to achieve pipeline-parallelism. 

\begin{table}[h] \small 
\centering
\caption{System Configuration.}
\label{tab:system}
\resizebox{.85\linewidth}{!}{
\begin{tabular}{r|c|c|c} 
 \hline
 & \multirow{2}{*}{GPU} & \multicolumn{2}{c}{FPGA} \\
                \hhline{~~|-|-}
                &&SpMM\textsuperscript{*} &win-attn\textsuperscript{$\S$}\\
 \hhline{=|=|==}
Number &  2 & \multicolumn{2}{c}{3} \\
Model & MI210 & \multicolumn{2}{c}{U280} \\
Process Node & 6 nm& \multicolumn{2}{c}{16nm} \\
Local Memory & 64GB HBM2e & \multicolumn{2}{c}{8GB HBM2}\\
Dynamic Power &300W &  55W & 50.2W\\ 
Static Power\textsuperscript{+} & 45W &  \multicolumn{2}{c}{19.5W}\\ 
 \hline
\end{tabular}
}

\footnotesize $\ast$ for our customized Sextans SpMM \quad + measured at complete idle state \\ $\S$ window-attention in BigBird settings
\end{table}

\noindent
\textbf{Implementation and Evaluation.}
%For CPU performance, we used the AMD AOCL library (AOCL-BLAS for GEMM and AOCL-Sparse for SpMM) on the 7v13 64-core CPUs available in our cluster.
We use FP32 precision for both device types. We design and implement a customized version of the Sextans architecture~\cite{song2022sextans} for SpMM on FPGA. The original Sextans design supports general SpMM in the form of $C = \alpha A \times B + \beta C$. We removed $\alpha$ and $\beta C$ that are unnecessary in our SpMM computation. The freed resources are used to increase the total number of computational units. The performance of FPGA GEMM is obtained from the state-of-the-art implementation~\cite{gemmfpga}. We use \texttt{rocsparse\_spmm} and \texttt{rocblas\_sgemm} for SpMM and GEMM kernels on GPU with ROCm 5.4.2 (the most stable version available when this work was started). 
Newer versions, such as ROCm 6.2.4, are available now~\cite{amdrocm}. 
We use PyTorch to implement the transformer models on GPUs, and we implement the FPGA-based sliding-window attention according to \cite{bai2024swat} using the BigBird settings as detailed in section~\ref{sec:sliding-window}.

The system's throughput and energy efficiency are measured by applying the schedule provided by {\name} on our current hardware build of heterogeneous system with PCIe\textsuperscript{\textregistered}~4.0 interconnect. Acknowledging the advancing trend of high-speed system-level connections, we extended our evaluation to include projections with the theoretical bandwidths of PCIe\textsuperscript{\textregistered}~5.0~\cite{pci} 
and CXL\textsuperscript{\texttrademark}~3.0~\cite{cxl}. 
Each kernel execution and data transfer is timed. Only the data transfer time is projected when considering other interconnect technologies. 
We use the corresponding XRT and ROCm libraries to measure the device powers for the system.

\noindent\textbf{{\name} Scheduling Objectives.}  
To assess the performance and energy efficiency of {\name} under various scheduling objectives and to explore the trade-offs, we experimented with the following settings.
The \textit{performance-optimized} mode prioritizes maximum system throughput, disregarding energy consumption. Conversely, the \textit{energy-optimized} configuration aims for the highest energy efficiency without throughput considerations. The \textit{balanced} setup seeks an optimal compromise, offering the most energy-efficient design while ensuring throughput remains at least 70\% of that in the \textit{performance-optimized} configuration.

\subsection{Accuracy of Estimation Model}
The accuracy of our estimation model is critical for the scheduler's ability to identify optimal schedules. While inaccuracies may lead to sub-optimal scheduling decisions, the scheduler is able to tolerate a certain level of error while still producing the optimal schedule or sub-optimal schedules without significantly compromising the overall system performance.
To assess the impact of estimation errors on scheduling, we evaluate the number of cases where the scheduler fails to identify the optimal schedule (\# sub-optimal). This is achieved by running the scheduler with the actual measured performance of the kernels and comparing the outcomes to the optimal schedules determined with the estimation model.

The relative loss (in \%) of the sub-optimal cases is assessed as the percentage reduction in system performance when using the sub-optimal schedule. For objectives optimized for throughput, the relative error is quantified as the performance loss. For energy-optimized objectives, it is measured as the loss of energy efficiency. We couldn't measure the loss in balance mode, where there's no singular 'optimal' schedule but rather an energy-performance trade-off.
Despite occasional discrepancies, the evaluation results demonstrate that the {\name} scheduler effectively identifies the most suitable schedule in the majority of cases. Even when it does not achieve the optimal outcome, the actual performance loss typically remains within acceptable limits. 

\begin{table}[h]
    \centering
\caption{Accuracy of {\name} scheduler on GNN workloads. throughput or energy loss is averaged on the sub-optimal cases.}
\label{tab:accuracy}
    \resizebox{\linewidth}{!}{
    \begin{tabular}{|c|c|c|c|}
         \hline
          \multicolumn{2}{|c|}{throughput-optimized}     & \multicolumn{2}{c|}{energy-optimized}  \\
         \hline
         \# sub-optimal & througput loss(\%)  & \# sub-optimal & energy loss(\%) \\
         \hline
         3/42 & 5.94\%      &4/42 & 2.46\% \\
    \hline
    \end{tabular}
    }
\end{table}

\subsection{{\name} Evaluation Results }
\label{sec:dype-eval}

% \subsubsection{{\name} enhances system throughput while conserving energy.}
\subsubsection{Performance and Energy Efficiency.}
Table~\ref{tab:dype-GNN} shows comparative analyses of {\name} throughput and energy efficiency (number of inferences per Joule) across three distinct scheduling modes set against various baselines for GNN workloads and sliding-window-based transformer models.
The improvement is calculated by the average across all interconnect technologies. For GNN models, it is averaged across all graph datasets. For transformer models, it is averaged across all combinations of lengths of the input sequence and window width for transformer models as depicted in Section~\ref{sec:sliding-window}.

\begin{table}[h]\small
    \centering
    \caption{Throughput (thp) and energy efficiency (eng) improvement of {\name} on GNN and transformers workloads.}
    \label{tab:dype-GNN}
    \resizebox{\linewidth}{!}{
    \begin{tabular}{|c|c|c|c|c|c|c|}
            \hline
          \multirow{2}{*}{Compared With}    & \multicolumn{2}{c|}{\name-perf-opt.} & \multicolumn{2}{c|}{\name-balanced}  & \multicolumn{2}{c|}{\name-energy-opt.} \\
          \hhline{|~|-|-|-|-|-|-|}
                            &thp. &eng.  &thp. &eng.  &thp. &eng.   \\
        \hline
        \multicolumn{7}{|c|}{GNN workloads} \\
        \hline
         FleetRec$^*$    & 2.06x  & 1.23x      & 1.95x & 1.42x      & 1.39x & 1.60x \\
         static      & 2.24x  & 1.35x      & 2.08x & 1.54x      & 1.39x & 1.68x \\
         theoretical-additive & 1.30x  & 1.53x      & 1.18x & 1.71x      & 0.79x     & 1.84x \\
         \hline
         FPGA-only   & 5.71x  & 1.88x      & 5.32x & 2.09x      & 3.40x & 2.23x  \\
         GPU-only    & 1.68x  & 1.17x      & 1.52x & 1.33x      & 1.03x & 1.45x \\
         \hline
         \multicolumn{7}{|c|}{Transformer workloads} \\
        \hline
        static/FleetRec$^*$     &1.18x&1.00x&1.11x&1.05x &0.72x&1.08x\\
        theoretical-additive    &1.33x&2.27x&1.28x&2.42x &0.89x&2.46x\\
        \hline
        FPGA-only      &9.26x&2.54x&9.17x&2.76x&6.49x &2.79x\\
        GPU-only       &1.28x&1.99x&1.21x&2.07x &0.77x&2.13x\\
        \hline
        \multicolumn{7}{|c|}{Average} \\
        \hline
        FleetRec$^*$ &1.53x, &1.09x, &1.45x, &1.20x, &0.99x, &1.29x\\
        Theoretical-additive &1.32x, &1.97x, &1.24x, &2.14x, &0.85x, &2.21x \\
        GPU-only $^*$ &1.44x, &1.66x, &1.33x, &1.77x, &0.87x, &1.86x\\
        \hline
    \end{tabular}
    }
\end{table}

\begin{table*}[h]
\centering
\caption{Scheduling Result of {\name} on GNN workloads.}
\resizebox{\textwidth}{!}{
\begin{tabular}{|l|lll|lll|lll|}
\hline
       & \multicolumn{3}{c|}{{PCIe\textsuperscript{\textregistered}}4.0} & \multicolumn{3}{c|}{{PCIe\textsuperscript{\textregistered}}5.0} & \multicolumn{3}{c|}{CXL\textsuperscript{\texttrademark}3.0} \\ \cline{2-10}
       & \multicolumn{1}{l|}{perf.-opt.} & \multicolumn{1}{l|}{balanced} & energy-opt. & \multicolumn{1}{l|}{perf.-opt.} & \multicolumn{1}{l|}{balanced} & energy-opt. & \multicolumn{1}{l|}{perf.-opt.} & \multicolumn{1}{l|}{balanced} & energy-opt. \\ \hline
GCN-OA & \multicolumn{1}{l|}{3F2G}       & \multicolumn{1}{l|}{3F1G}     & 2F1G        & \multicolumn{1}{l|}{3F2G}       & \multicolumn{1}{l|}{3F2G}     & 2F1G1F1G    & \multicolumn{1}{l|}{2F2G}       & \multicolumn{1}{l|}{2F1G1F1G} & 2F1G1F1G    \\ \hline
GCN-OP & \multicolumn{1}{l|}{3F2G}       & \multicolumn{1}{l|}{3F2G}     & 2F1G        & \multicolumn{1}{l|}{3F2G}       & \multicolumn{1}{l|}{3F2G}     & 2F1G        & \multicolumn{1}{l|}{3F2G}       & \multicolumn{1}{l|}{3F2G}     & 2F1G        \\ \hline
GCN-S1 & \multicolumn{1}{l|}{2G}         & \multicolumn{1}{l|}{3F1G}     & 3F1G        & \multicolumn{1}{l|}{2G}         & \multicolumn{1}{l|}{2G}       & 3F1G        & \multicolumn{1}{l|}{2G}         & \multicolumn{1}{l|}{2G}       & 3F1G        \\ \hline
GCN-S2 & \multicolumn{1}{l|}{2G}         & \multicolumn{1}{l|}{3F2G}     & 2F1G        & \multicolumn{1}{l|}{2G}         & \multicolumn{1}{l|}{3F1G}     & 3F1G        & \multicolumn{1}{l|}{2G}         & \multicolumn{1}{l|}{2G}       & 3F1G        \\ \hline
GCN-S3 & \multicolumn{1}{l|}{2G}         & \multicolumn{1}{l|}{2G}       & 1G          & \multicolumn{1}{l|}{3F2G}       & \multicolumn{1}{l|}{3F1G}     & 2F1G        & \multicolumn{1}{l|}{3F2G}       & \multicolumn{1}{l|}{3F1G}     & 2F1G        \\ \hline
GCN-S4 & \multicolumn{1}{l|}{3F2G}       & \multicolumn{1}{l|}{2G}       & 1G          & \multicolumn{1}{l|}{2F2G}       & \multicolumn{1}{l|}{2G}       & 1G          & \multicolumn{1}{l|}{2F2G}       & \multicolumn{1}{l|}{2G}       & 1G          \\ \hline
GIN-OA & \multicolumn{1}{l|}{3F2G}       & \multicolumn{1}{l|}{3F2G}     & 2F1G        & \multicolumn{1}{l|}{3F2G}       & \multicolumn{1}{l|}{3F2G}     & 2F1G1F1G    & \multicolumn{1}{l|}{2F2G}       & \multicolumn{1}{l|}{1F1G1F1G} & 1F1G1F1G    \\ \hline
GIN-OP & \multicolumn{1}{l|}{3F2G}       & \multicolumn{1}{l|}{3F2G}     & 1G          & \multicolumn{1}{l|}{2F2G}       & \multicolumn{1}{l|}{2F2G}     & 1F1G2F1G    & \multicolumn{1}{l|}{3F2G}       & \multicolumn{1}{l|}{2F2G}     & 1F1G2F1G    \\ \hline
GIN-S1 & \multicolumn{1}{l|}{2G}         & \multicolumn{1}{l|}{3F1G}     & 3F1G        & \multicolumn{1}{l|}{2G}         & \multicolumn{1}{l|}{2G}       & 3F1G        & \multicolumn{1}{l|}{2G}         & \multicolumn{1}{l|}{2G}       & 3F1G        \\ \hline
GIN-S2 & \multicolumn{1}{l|}{2G}         & \multicolumn{1}{l|}{3F1G}     & 2F1G        & \multicolumn{1}{l|}{2G}         & \multicolumn{1}{l|}{3F1G}     & 3F1G        & \multicolumn{1}{l|}{2G}         & \multicolumn{1}{l|}{2G}       & 3F1G        \\ \hline
GIN-S3 & \multicolumn{1}{l|}{2G}         & \multicolumn{1}{l|}{2G}       & 1G          & \multicolumn{1}{l|}{3F2G}       & \multicolumn{1}{l|}{3F1G}     & 2F1G        & \multicolumn{1}{l|}{3F2G}       & \multicolumn{1}{l|}{3F2G}     & 2F1G        \\ \hline
GIN-S4 & \multicolumn{1}{l|}{2F2G}       & \multicolumn{1}{l|}{2G}       & 1G          & \multicolumn{1}{l|}{2F2G}       & \multicolumn{1}{l|}{2G}       & 1G          & \multicolumn{1}{l|}{2F2G}       & \multicolumn{1}{l|}{2G}       & 1G          \\ \hline
\end{tabular}
}
\label{tab:all-schedules}
\end{table*}

\textbf{{\name}'s dynamicity improves the overall performance.} 
For GNN workloads, {\name} consistently achieves higher throughput and energy efficiency than both the \textit{static} and \textit{FleetRec} baselines, primarily due to its dynamic scheduling capabilities. This dynamicity allows {\name} to adaptively align computing resources with the varying demands of different workloads and system conditions, optimizing both performance and energy usage. 

To better show the adaptability of {\name}, Table~\ref{tab:all-schedules} lists all the optimal schedules with respect to specific datasets and interconnect technologies. Each schedule is represented by a mnemonic, such as 3F2G, which denotes a pipeline configuration where the first stage is allocated to 3 FPGAs and the second to 2 GPUs. The table clearly shows that the optimal scheduling strategy can vary significantly depending on the workload characteristics, available data transfer bandwidth, and specific design objectives. We further count the configurations where the best schedule can be found by the \textit{static} or \textit{FleetRec} in the table, which represent only a small portion of the schedules: 8 out of 108 cases. This stark contrast highlights the substantial benefits of dynamic scheduling in heterogeneous computing environments, particularly when the computational demands and data characteristics can change dramatically. 

The improvements achieved by {\name} are less pronounced in transformer workloads. This is because, in transformer architectures, only the attention computations are suitable for FPGA acceleration, which significantly narrows the design space available for optimization. Given that each layer of a transformer involves just one attention computation, and each of these computations bears the same computational load across all layers, the \textit{FleetRec} approach effectively becomes indistinguishable from the \textit{static} method in this context. Despite these constraints, {\name} still manages to deliver enhancements in both throughput and energy efficiency, particularly in performance-optimized and balanced modes. In performance-optimized mode, {\name} demonstrates a clear throughput improvement. In balanced mode, the framework not only boosts throughput but also enhances energy efficiency, showcasing its ability to adaptively balance performance and power consumption to meet dual objectives. However, in energy-optimized mode, the results indicate a trade-off where a slight increase in energy efficiency comes at the cost of considerable throughput degradation, suggesting that pursuing an aggressive energy efficiency strategy may not be suitable for transformer workloads under the current system configuration and workload characteristics. 

\begin{figure*}[t]
    \centering
    \includegraphics[width=\textwidth]{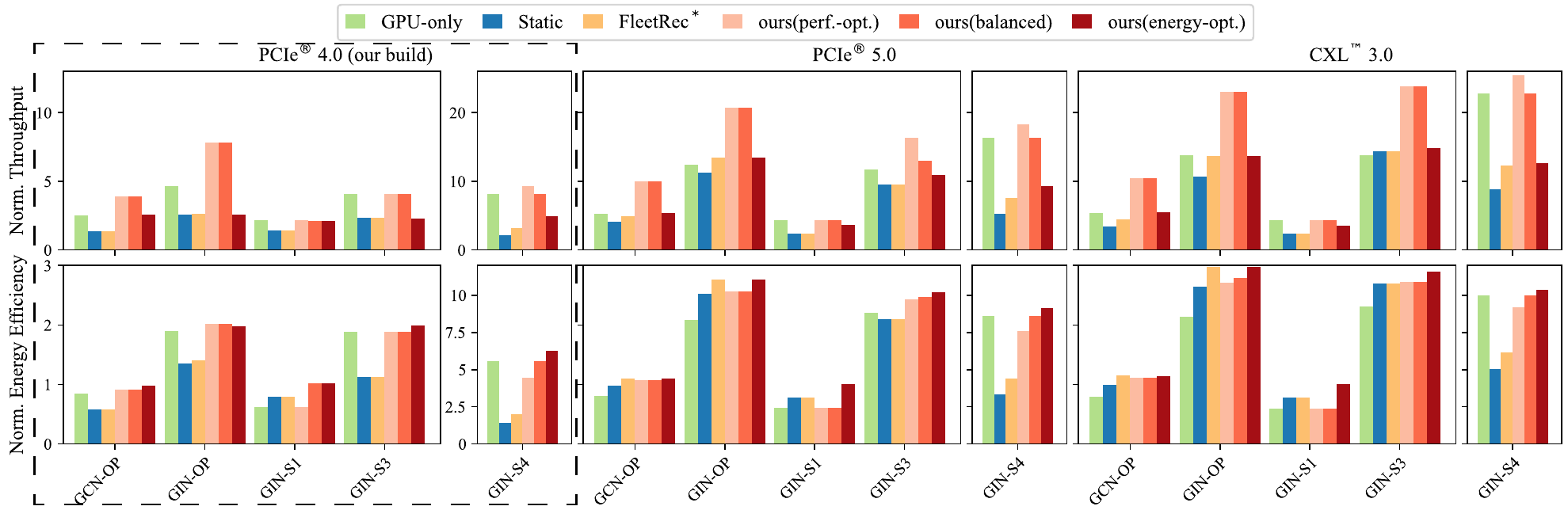}
    \caption{Througput and energy efficiency comparison between {\name} and the baselines, normalized to FPGA-only.}
    \label{fig:GNN-perf-energy}
    \vspace{-10pt}
\end{figure*}

\textbf{Improvements with heterogenity.}
When evaluated against \textit{GPU-only} and \textit{FPGA-only} system settings, {\name} demonstrates a distinct performance advantage. 
Specifically, it outperforms the \textit{GPU-only} configuration, leveraging the superior efficiency of FPGAs in handling sparse matrix operations on SpMM. When compared to \textit{theoretical-additive} setting, where the throughput of \textit{GPU-only} and \textit{FPGA-only} systems is aggregated, {\name} still exhibits better throughput and energy efficiency. This outcome underscores the benefits of leveraging heterogeneous devices optimally—each device is utilized for computations where it performs best, achieving an outcome where "one plus one equals more than two."

\subsubsection{Impact of Workloads Characteristics and System Configurations}
{\name}'s performance enhancements show significant variance based on the specific characteristics of the workload. For GNN workloads, this variance largely stems from factors such as dataset sparsity and feature length, which affect the arithmetic intensity of SpMM kernels. Notably, in cases like datasets S1 and S4, arithmetic intensity can vary by two orders of magnitude, leading to differing suitabilities for GPU and FPGA processing.

Figures~\ref{fig:GNN-perf-energy} illustrate the throughput and energy efficiency improvements of different approaches on GNN datasets, normalized to the \textit{FPGA-only} setup. We selected five workloads to demonstrate the impact of workload characteristics and interconnect technology.

Comparing \textit{Static}, FleetRec\cite{jiang2021fleetrec}, and {\name}, we observe that FleetRec consistently outperforms or matches \textit{static} thanks to the extra flexibility of adjusting the number of devices for each pipeline stage. Furthermore, {\name} consistently outperforms both by allowing kernel allocation between both types of devices. 

Regarding model characteristics, all approaches show better throughput and energy efficiency relative to \textit{FPGA-only} in GIN models compared to GCN models. Two such instances are GCN-OP and GIN-OP, as shown in the Figure~\ref{fig:GNN-perf-energy}. This is attributed to GIN's higher number of invokes to the GeMM kernel, resulting in a higher dense-sparse ratio that is less favorable for a pure FPGA setup. However, although GIN models contain a greater ratio of dense kernels that are more favorable for GPUs, this does not limit the heterogeneous setting to improve the overall performance: {\name}'s improvement over GPU-only is similar on GCN-OP and GIN-OP.

Data characteristics also play a crucial role. GIN-S1, GIN-S3, and GIN-S4 demonstrate distinct trends as interconnect bandwidth increases: while all approaches show comparable benefits from improved interconnect technologies for GIN-S1 and GIN-S4, {\name} exhibits significantly higher gains on GIN-S3 compared to other approaches. These differences stem from varying input sparsities and shapes, as listed in Table~\ref{tab:dataset}, leading to different stage times and balance between sparse and dense computations. 

For GIN-S1, the graph's high number of edges and low sparsity make the SpMM kernel less advantageous for FPGAs. 
%In a heterogeneous schedule, the SpMM stage becomes dominating, even when all three FPGAs are used, and the GPUs are mostly idle after computing the GeMM kernels, leading to an imbalanced pipeline.
Conversely, GIN-S4's high sparsity makes GeMM kernels the bottleneck stage, favoring heterogeneous system schedules where the SpMM stage can be hidden through pipelining. In these extreme cases, the dense-sparse computation ratio strongly influences schedule preferences, with minimal impact from interconnect technology. GIN-S3, with intermediate sparsity, exhibits balanced SpMM and GeMM stage times. Here, interconnect technology significantly impacts scheduler choice: higher bandwidth reduces FPGA communication overhead, improving heterogeneous scheduling performance. In conclusion, the diverse scheduling preferences observed across different datasets underscore the necessity for dynamic schedulers capable of adapting to varying workload characteristics.

\begin{figure}[h]
    \centering
    \includegraphics[width=\linewidth]{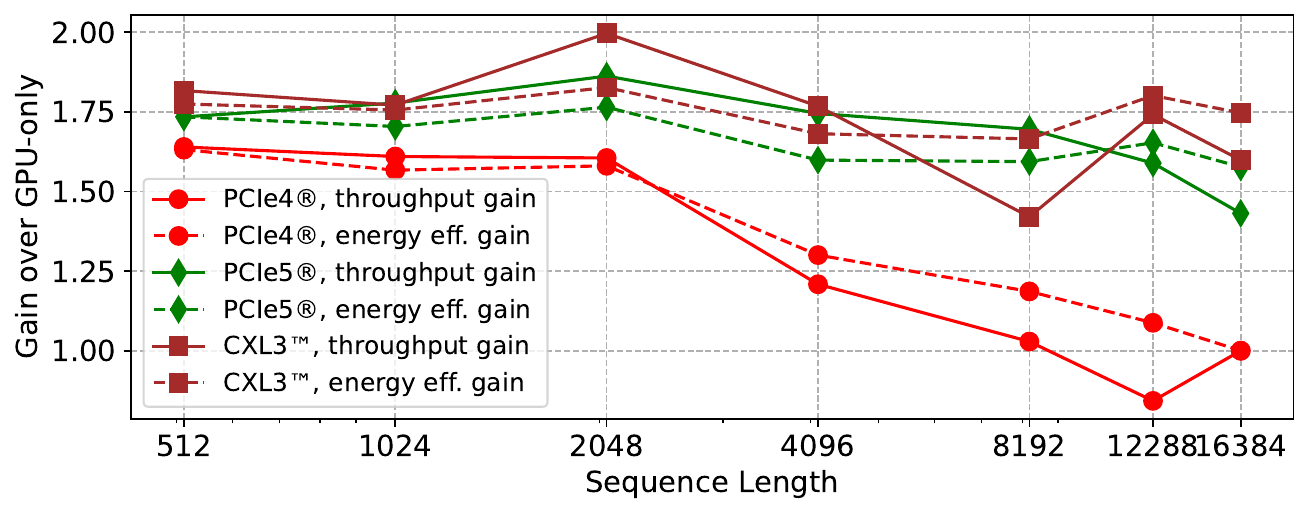}
    \caption{Throughput and energy efficiency gain of {\name} over GPU-only on sliding-window-bases transformers workloads of window size fixed to 512.}
    \label{fig:dype-transformers-comm}
    \vspace{-10pt}
\end{figure}

Observed with GNN workloads, we note that as sparsity in the dataset increases (GIN-S1/S3/S4), the optimal scheduling increasingly favors FPGA inclusion to better utilize the system's computation resources. However, this strategy does not extend to transformer workloads. 
As shown in Figure~\ref{fig:dype-transformers-comm} (e.g. PCIe\textsuperscript{\textregistered} 4.0), when the sparsity increases along with the input sequence, the advantage of {\name} is less significant compared to the GPU-only baseline.
Although sparsity in transformers also increases with sequence length, the accompanying rise in communication overhead surpasses the advantages derived from heterogeneity. This imbalance negates the benefits of additional computational resources. As there is no common rule for all workloads, apart from data characteristics, it is also crucial to tailor scheduling decisions for each specific application within heterogeneous systems.

\begin{figure}
    \centering
    \includegraphics[width=0.8\linewidth]{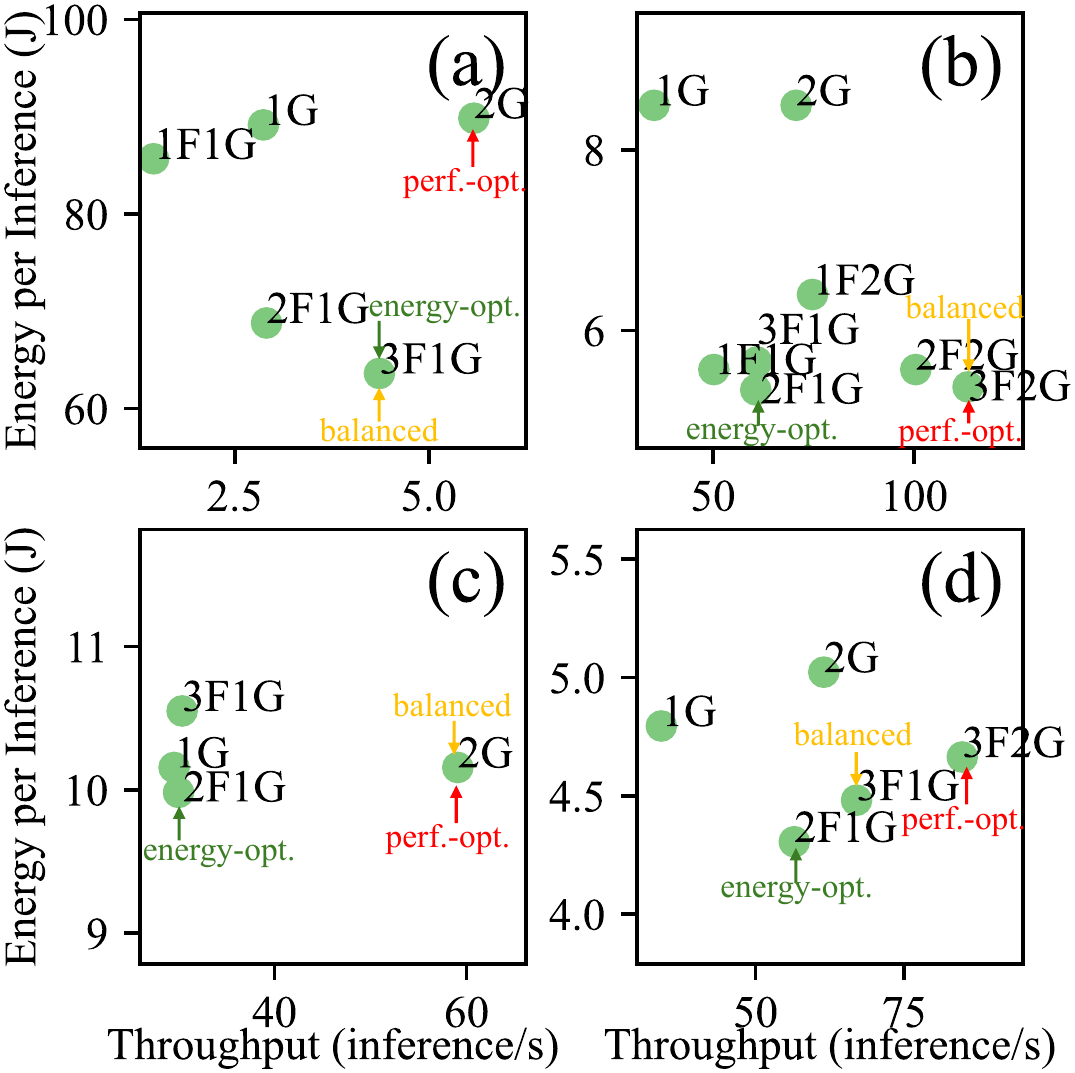}
    \caption{Design space exploration on GCN-S1 and GCN-OP, evaluated on system with PCIe\textsuperscript{\textregistered}4.0 interconnect in balanced mode. Only Pareto-optimal schedules are shown in terms of throughput, energy, and device number. Bottom right corner is the best. (a) GCN, Syn1; (b) Sliding-Window-Based Transformers, input length 2048, window width 512; (c) Sliding-Window-Based Transformers, input length 12288, window width 2048; (d) GCN, ogbn-arxiv;}
     \label{fig:design-space-exploration}
\end{figure}

\subsubsection{Importance of Configurable Performance-Energy Trade-off}
Unlike previous work that primarily focuses on maximizing performance, {\name} offers a configurable performance-energy trade-off. This flexibility has proven crucial in our experiments, as we found that significant gains in performance or energy efficiency often come at the expense of the other.
As shown in Table~\ref{tab:dype-GNN}, the \textit{energy-optimized} mode improves the energy efficiency with significant throughput loss. Some concrete examples can be found in Figure~\ref{fig:design-space-exploration}. 

Figure~\ref{fig:design-space-exploration}a shows how energy efficiency can be considerably enhanced at the cost of only a minor performance loss, a case where the energy-optimized mode would perform well. Conversely, Figure~\ref{fig:design-space-exploration}b-c depict scenarios where the energy-optimized model causes significant performance loss.  Figure~\ref{fig:design-space-exploration}d presents a case where a third Parento-optimal case is in between the energy-optimized one and the performance-optimized one.

These observations underscore the importance of the \textit{balanced} mode in {\name}. This mode ensures a minimal acceptable level of throughput, which effectively constrains the design exploration to a reasonable range, preventing extreme trade-offs that could render the system impractical for its intended applications.

\section{\zhen{Related works}}
% \zhen{
% \noindent\textbf{FPGAs in heterogeneous systems.} The utilization of FPGAs in heterogeneous systems for handling irregularities has been extensively explored to improve performance and energy efficiency across various domains. FPGAs are employed for data management tasks such as managing memory accesses between GPUs and main memory~\cite{li2022hyperscale}, and as smart Network Interface Cards (NICs) that preprocess and distribute data from networks to accelerators~\cite{smart-nic-1, smart-nic-2, smart-nic-3, smart-nic-4}. In addition to data management, FPGAs have been integrated into CPU-FPGA heterogeneous systems for Graph Neural Network (GNN) training~\cite{zeng2020graphact, zhang2022low}, where CPUs handle memory operations and FPGAs perform computational tasks. FPGA-GPU combinations have been utilized to accelerate scientific simulations and convolutional neural networks, often employing static workload distribution strategies~\cite{heterogeneous-cluster-static-division, heterogeneous-static-division-1}. In embedded systems, FPGAs are combined with GPUs to improve performance and energy efficiency through static offloading of computations~\cite{heterogeneous-embbeded-simple-division-1, heterogeneous-embbeded-simple-division-2, heterogeneous-embbeded-divison-in-FPGA}. While these approaches demonstrate the benefits of FPGAs in heterogeneous systems, they often rely on static resource allocation.
% }

\noindent
\textbf{Scheduling in Heterogeneous Distributed Systems} has been addressed using various methods, including reinforcement learning and meta-heuristic algorithms. \cite{orhean2018new} propose a scheduling approach using reinforcement learning for heterogeneous distributed systems, assuming known task execution times to optimize scheduling decisions. Similarly, \cite{shirvani2020hybrid,karatza2001job} present heuristic-based algorithms for workflow scheduling in heterogeneous distributed computing systems, combining heuristic methods to find near-optimal solutions. \cite{vasile2015resource} introduce a system resource-aware hybrid scheduling algorithm that models hardware resources to enhance scheduling efficiency. In the context of distributed deep learning, HeterPS~\cite{liu2023heterps} employs reinforcement learning-based scheduling in heterogeneous environments, enabling resource-aware and pipeline-parallel execution. However, these approaches often assume a given workload with known performance profiles, either by assumption or based on prior measurements. In contrast, our work addresses scenarios where rescheduling occurs at runtime based on data characteristics. Additionally, these studies typically focus on scheduling for single inference tasks without considering the streaming of continuous inferences, which is common in machine learning workloads. 

\noindent
\textbf{ML Workloads Acceleration on Heterogeneous Edge Devices.}
In edge computing, substantial research has been conducted on the use of heterogeneous systems for offloading machine learning workloads. \cite{li2023adaptive, kang2017neurosurgeon, hu2019dynamic, jeong2018ionn} dynamically offload a part of the target workload to the edge server or/and to the cloud to minimize the latency of ML workloads on the terminal edge device. 
These works explore offloading computations to edge devices equipped with CPUs and GPUs to improve performance and reduce latency. However, they generally do not consider the sparsity of machine learning workloads and do not reschedule computations based on data characteristics.
While previous studies contribute valuable insights into scheduling and resource allocation in heterogeneous systems, they often rely on static assumptions and do not adapt to dynamic changes in workload characteristics. Our approach differs by dynamically adjusting resource allocations and schedules in response to real-time changes in data characteristics, particularly sparsity levels in computations.

\noindent
\textbf{Resource Allocation and Scheduling in Clusters.} Several works have proposed dynamic and flexible strategies for cluster resource allocation. ElasticFlow~\cite{elasticflow} and VirtualFlow~\cite{virtualflow} emphasize hardware virtualization, enabling dynamic resource allocation. Optimus~\cite{optimus} uses resource-performance models to determine allocations dynamically, akin to our approach. However, these methods rely on \textit{data-parallelism}, where the entire workload must be executed on a single device, with different devices handling separate batches or data inputs. Consequently, they do not fully exploit the heterogeneity within the workload.

% \noindent\textbf{Pipeline Parallelism in Heterogeneous Systems.} HetPipe \cite{park2020hetpipe} presents a heterogeneous GPU system designed for DNN training that incorporates pipeline parallelism. However, this system requires manual selection from pre-defined workload distribution strategies. AdaPipe \cite{adapipe} enhances this approach by focusing on balancing pipeline execution in DNN training, emphasizing the impact of recomputation strategies rather than device allocation. Muri~\cite{Muri} enables the parallelism between Storage IO operations, network Operations, CPU and GPU executions while focusing on the parallelism among different \zhen{compute kernels}.

\noindent

\section{Conclusion}
This paper addresses the challenge of dynamic workload management in heterogeneous environments. We introduced the  {\name} framework, which incorporates performance estimation models, adaptable system settings, and a dynamic scheduler. This scheduler adeptly navigates the design space, balancing data transfers and computations across heterogeneous devices to harness each's unique strengths on suitable parts of the workload.

Our findings underscore the potential of heterogeneous systems to handle the increasingly varied and complex demands of modern computing tasks, especially as data sparsity grows in fields like machine learning and graph processing. As such,  {\name} represents a critical step forward, laying the groundwork for further exploration into scalable, high-performance, and energy-efficient heterogeneous computing architectures.

% Add biographies here
\begin{IEEEbiography}[{\includegraphics[width=1in,height=1.25in,clip,keepaspectratio]{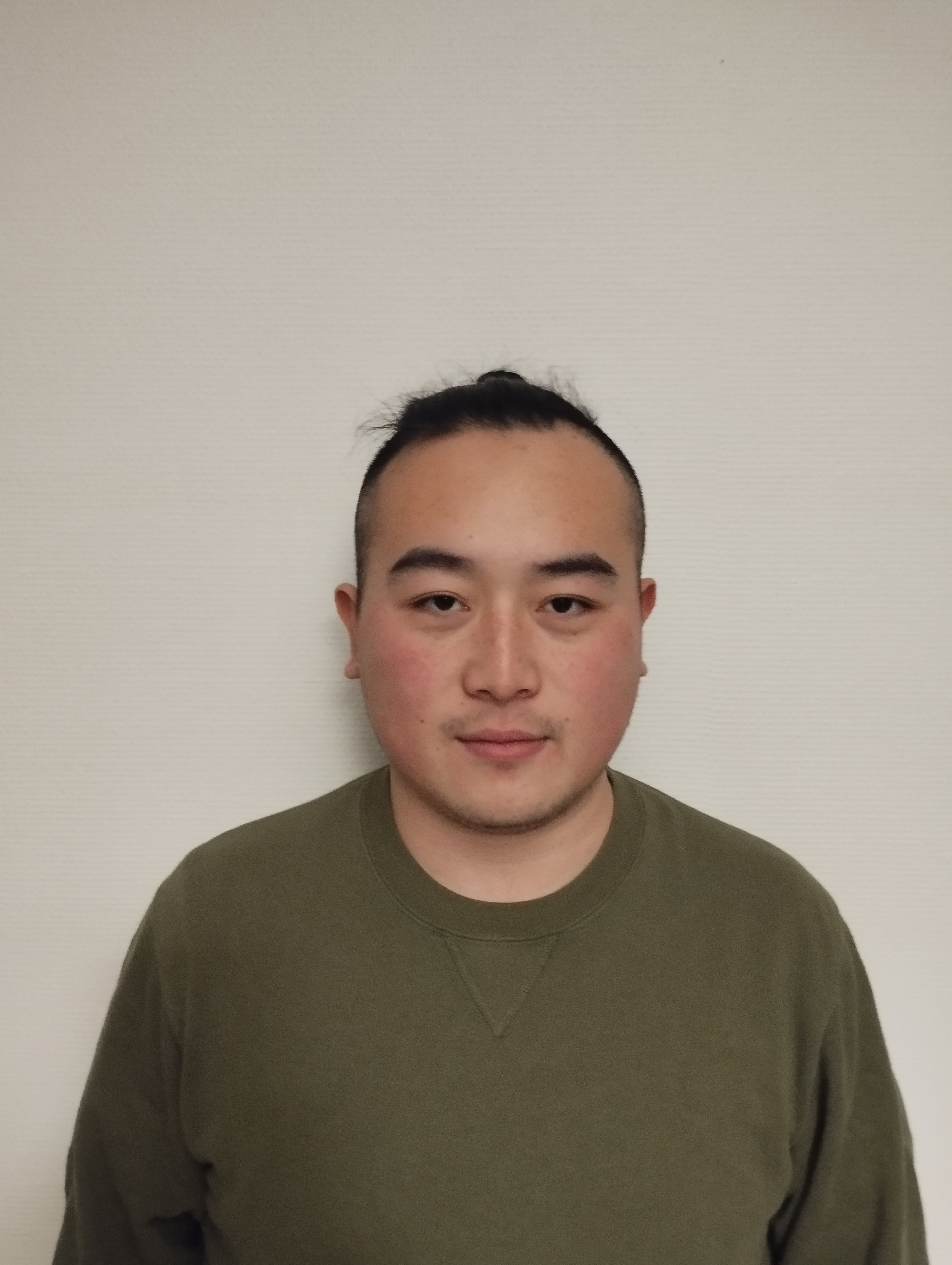}}]{Zhenyu Bai}
Zhenyu Bai is currently a Research Fellow in the Department of Computer Science, National University of Singapore. His research interests include transformer acceleration.
\end{IEEEbiography}

\begin{IEEEbiography}[{\includegraphics[width=1in,height=1.25in,clip,keepaspectratio]{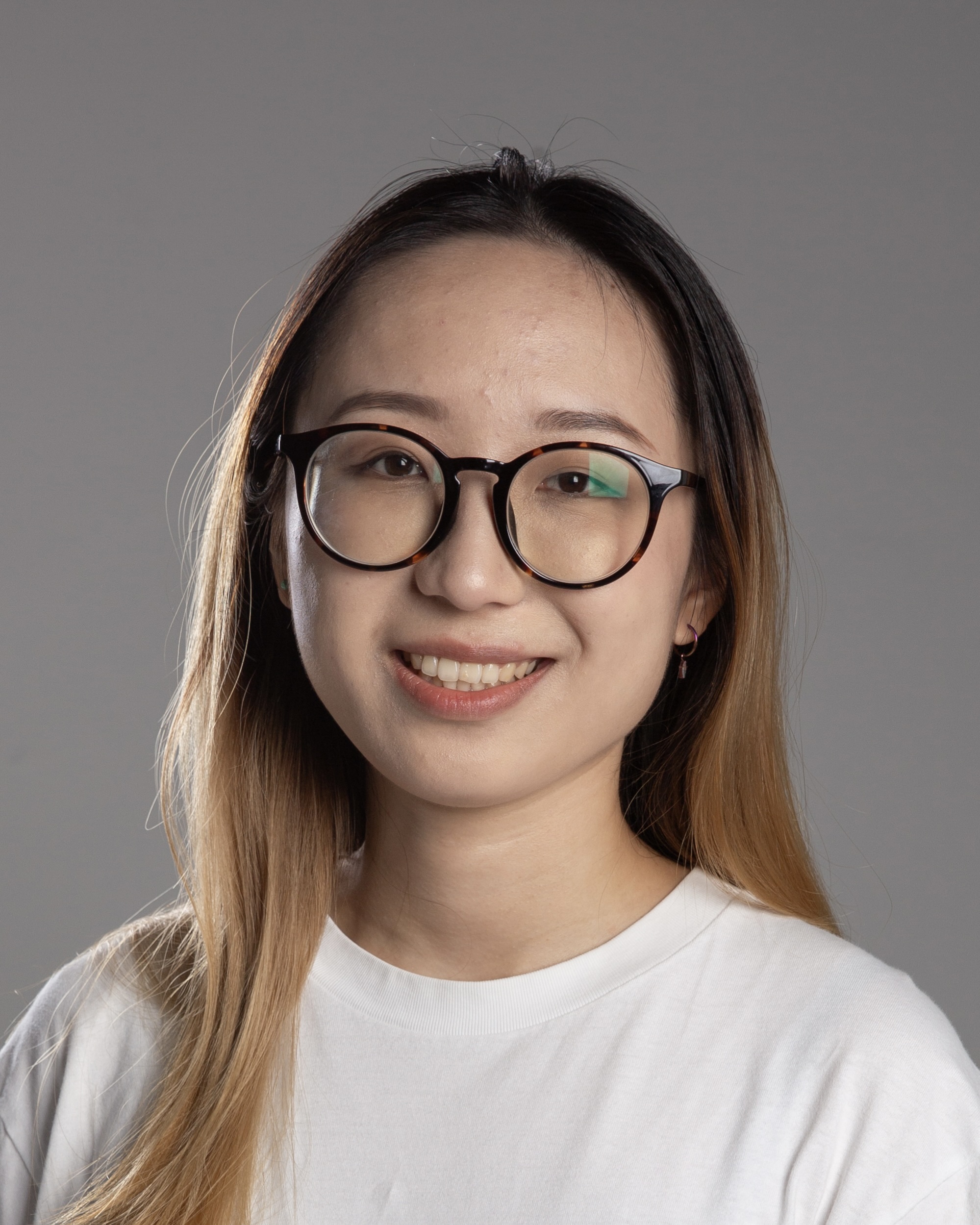}}]{Dan Wu}
Dan Wu is a Ph.D. candidate at the National University of Singapore. Her research focuses on software-hardware co-design and graph-related problem acceleration.
\end{IEEEbiography}

\begin{IEEEbiography}[{\includegraphics[width=1in,height=1.25in,clip,keepaspectratio]{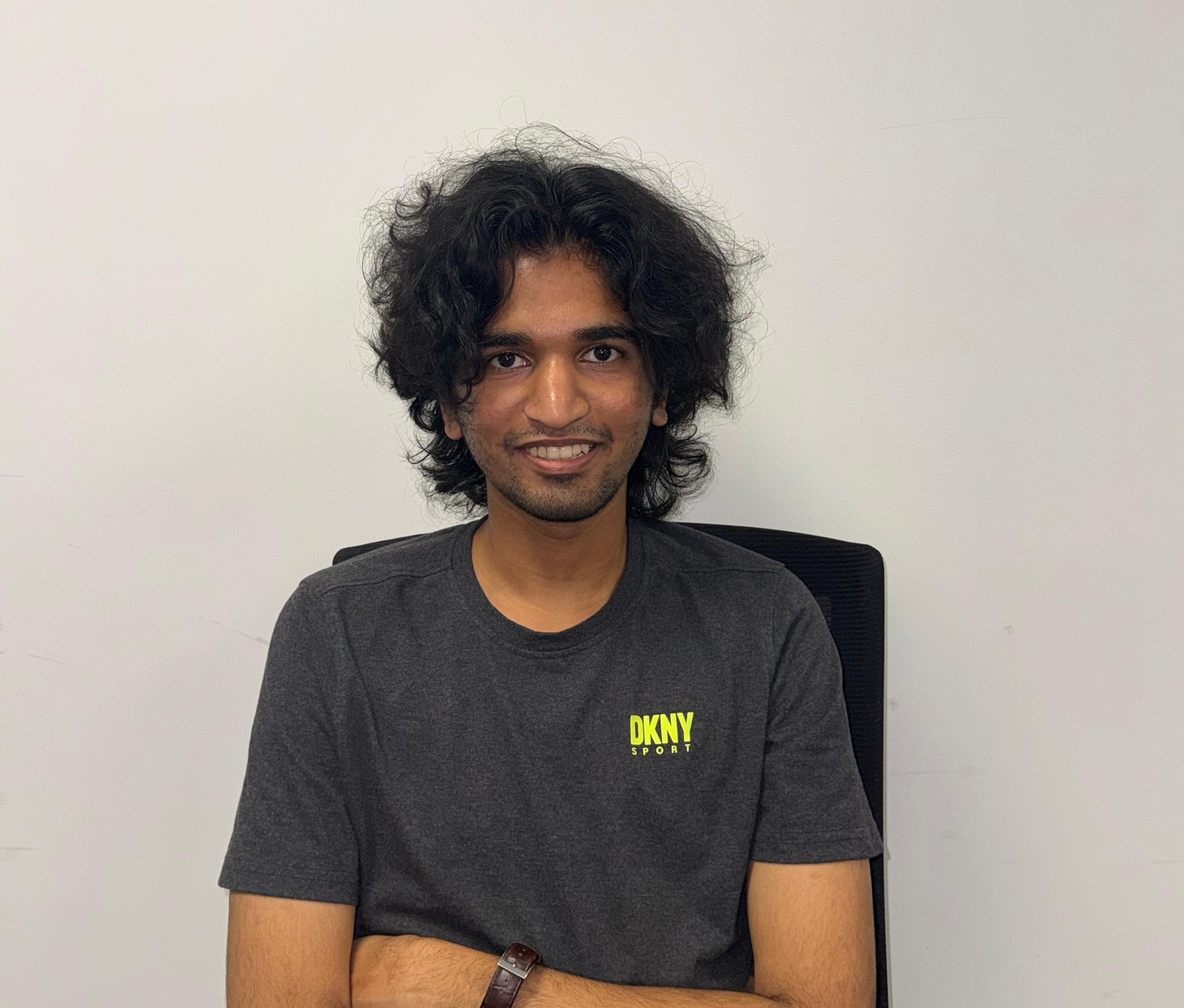}}]{Pranav Dangi}
Pranav Dangi is a Ph.D. student at National University of Singapore. His research interests include heterogeneous computing and reconfigurable architectures.
\end{IEEEbiography}

\begin{IEEEbiography}[{\includegraphics[width=1in,height=1.25in,clip,keepaspectratio]{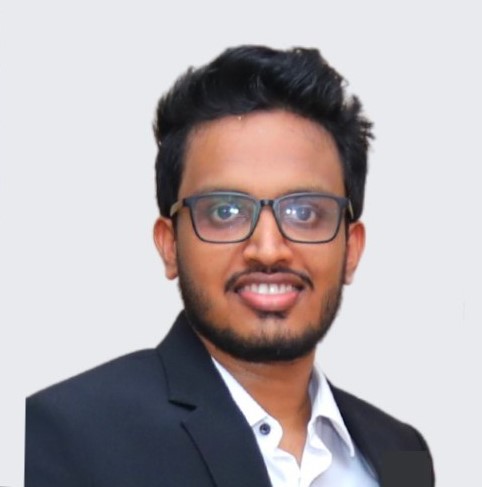}}]{Dhananjaya Wijerathne}
Dhananjaya Wijerathne received a Ph.D. in Computer Science from the National University of Singapore in 2023. He is currently a Member of Technical Staff at Advanced Micro Devices (Singapore) Pte. Ltd. 
\end{IEEEbiography}

\begin{IEEEbiography}[{\includegraphics[width=1in,height=1.25in,clip,keepaspectratio]{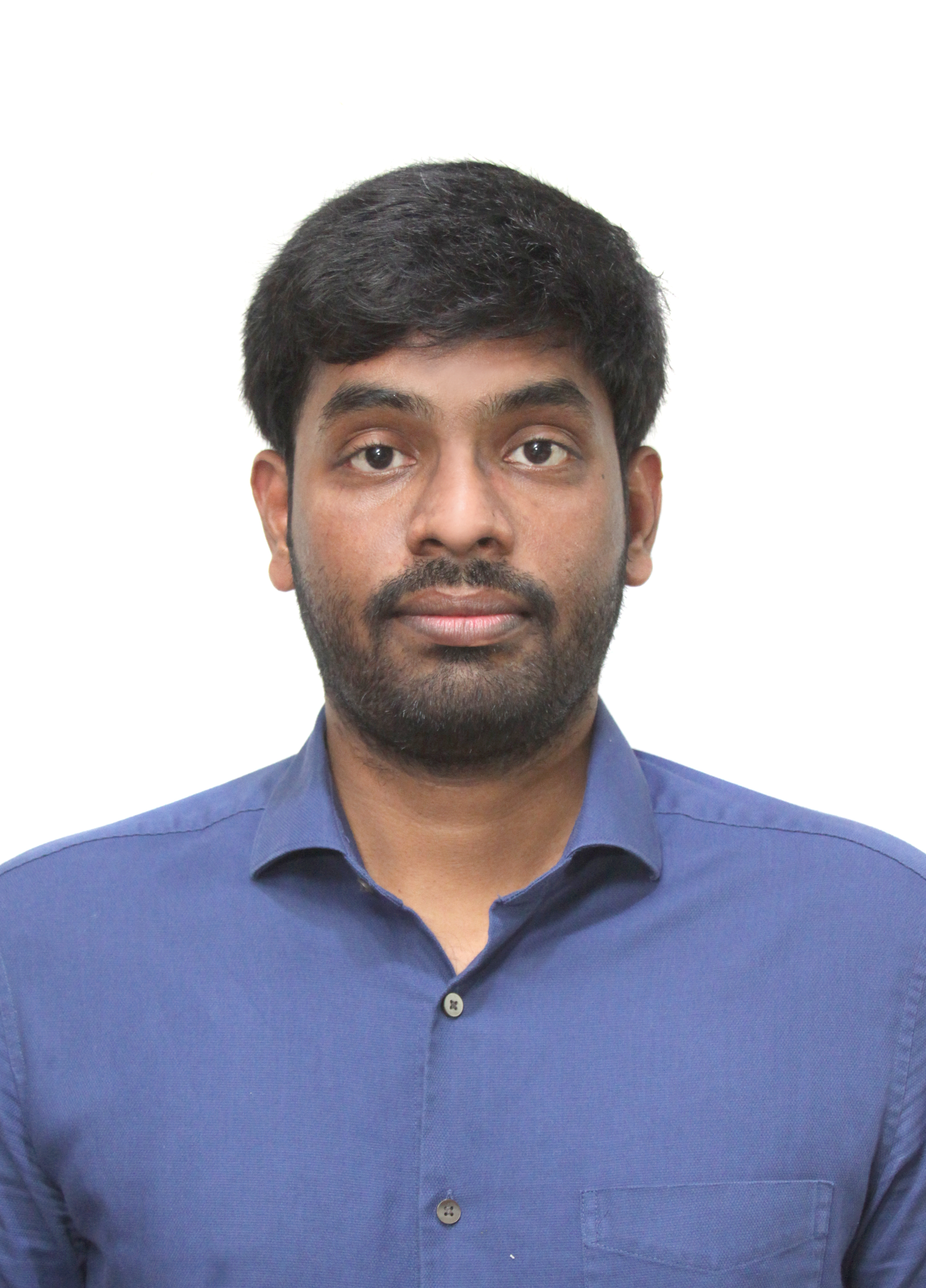}}]{Venkata Pavan Kumar Miriyala}
Venkata Pavan Kumar Miriyala received a Ph.D. in Electrical and Computer Engineering from the National University of Singapore (NUS), Singapore, in 2021. He is currently a Member of Technical Staff at Advanced Micro Devices (Singapore) Pte. Ltd. 
% He has also worked with Qualcomm India Pvt. Ltd., NUS, IBM-Research, Tokyo, Japan, and IIT Hyderabad, Hyderabad, India.
\end{IEEEbiography}

\begin{IEEEbiography}[{\includegraphics[width=1in,height=1.25in,clip,keepaspectratio]{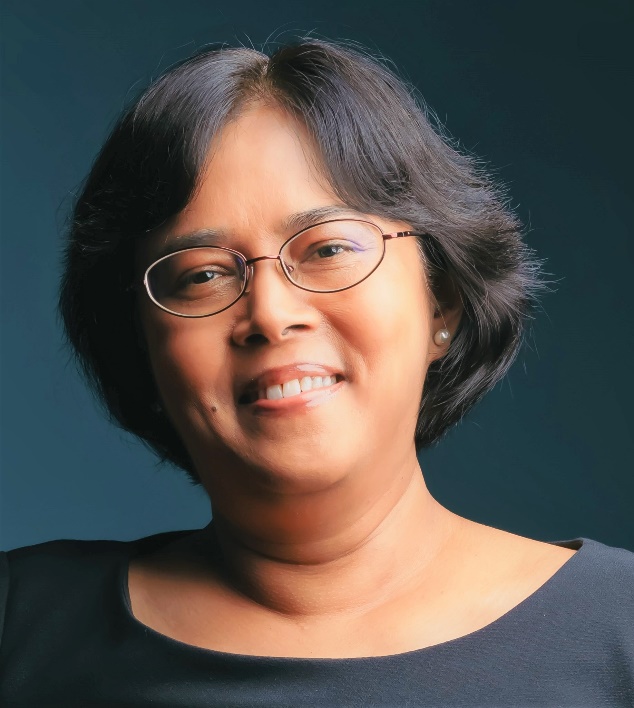}}]{Tulika Mitra}
Tulika Mitra is a Professor at the National University of Singapore. Her research interests include embedded systems, heterogeneous computing, and Coarse-Grained Reconfigurable Array.
\end{IEEEbiography}

\bibliographystyle{IEEEtran}
\bibliography{sample-base}

\end{document}

%% file: scheduling_new.tex
\subsection{\zhen{Scheduling Algorithm}}

\textsc{DyPe} employs a dynamic programming-based scheduling algorithm, detailed in Algorithm~\ref{alg:scheduling}, to strategically allocate computational kernels across available FPGAs and GPUs. While the algorithm is applicable to various device types, we focus on FPGAs and GPUs to reflect our prototype setup.

% The algorithm accepts as inputs the sequence of operations in the workload ($wl$) and the numbers of available FPGAs ($n_{\text{FPGA}}$) and GPUs ($n_{\text{GPU}}$). The performance estimation function $f_{\text{perf}}(K, \textit{dev}, n)$ predicts the execution time for a set of kernels $K$ when executed on $n$ devices of type \textit{dev}. The communication cost function $f_{\text{comm}}$ estimates data transfer time between pipeline stages and evaluates costs for both destination and source devices. The function $f_{\text{eng}}$ estimates the energy consumption of a given schedule.

The algorithm accepts the list of operations in the workload $wl$ and the counts of available FPGAs $n_{\text{F}}$ and GPUs $n_{\text{G}}$. It relies on three functions: performance estimation $f_{\text{perf}}$ predicts execution time for a group of kernels when executed on a specific number of devices of a particular device type. Communication cost estimation $f_{\text{comm}}$ estimates data transfer time between pipeline stages, assessing costs for destination and source devices. Energy estimation $f_{\text{eng}}$ computes energy consumption for a given schedule.

The algorithm builds a dynamic programming table $dp[i][f][g]$ representing the best scheduling option for the kernels $wl[0:i]$ with $f$ FPGAs and $g$ GPUs. Two tables, $dp_{\text{perf}}$ and $dp_{\text{eng}}$, track optimal schedules for performance and energy, respectively. These tables are updated independently, with performance optimization strategies highlighted in blue and energy optimization-related operations in orange, as shown in Algorithm~\ref{alg:scheduling}. Each entry in the $dp$ table stores the execution time of each stage and device assignments.

% We construct a dynamic programming table $dp[i][f][g]$, which records the best schedule after the first $i$ kernels have been scheduled using $f$ FPGAs and $g$ GPUs. The tables $dp_{\text{perf}}$ and $dp_{\text{eng}}$ record the best schedules optimized for performance and energy separately, respectively. These tables are updated separately, and operations related to energy optimization are marked in orange in Algorithm~\ref{alg:scheduling}. Each schedule in the $dp$ table contains information about the execution time of each stage and the device allocation. 

The table is filled iteratively through nested loops over $i$, $f$, and $g$ in lines 6–8 of Algorithm~\ref{alg:scheduling}. For each kernel $i$, the algorithm explores scheduling options by considering two strategies: \textbf{1) multiple devices can be allocated to the same kernel}, and \textbf{2) multiple consecutive kernels can be grouped together to form one pipeline stage, with these kernels executed sequentially by the same devices.} To implement these strategies, the algorithm references previous states in the dynamic programming table. Specifically, it looks back to $dp[i $-$ j][f$-$n_f][g]$ to consider grouping kernels from $i $-$ j$ to $i$ into a new pipeline stage using $n_f$ FPGAs, or to $dp[i $-$ j][f][g $-$ n_g]$ when using $n_g$ GPUs. 
% The complexity of the algorithm can be deduced by following the nested loops: $\mathcal{O}(|wl|^2 \times n_{FPGA}^2 \times n_{GPU}^2)$.

Without loss of generality, we detail the case where the stage accommodating the $i^\text{th}$ kernel is allocated to FPGA(s) ($n_f > 0$, $n_g $=$ 0$). In this scenario, we first retrieve information about the last stage of the previous schedule (lines 10–15) to accurately account for data transfer time to the destination devices, i.e., FPGAs (line 17). In the case of energy optimization, the $prev\_stage$ is obtained from the $dp_{\text{eng}}$ table. We then calculate the execution time of the new stage containing kernel $i$ by summing the kernel execution time and the incoming data transfer time (line 19). Additionally, we include the data transfer cost to the last stage of the previous schedule when adding the new stage (line 21). The longest stage duration in the updated pipeline is determined by taking the maximum among: 1) the previous stage's duration with updated communication cost, 2) the longest stage duration recorded thus far, and 3) the duration of the new stage. The obtained new pipeline time is compared and used to update the best schedule in the $dp$ table (lines 25–27).

% \zhen{For energy optimization, a new pipeline stage may become the longest stage of the pipeline, potentially introducing more idleness to the previously scheduled stages. Therefore, the energy evaluation must account for the entire pipeline. We create the new schedule from the previous one in line 31 and calculate its energy consumption based on the idleness in each stage, the data transfers, and the kernel execution. The power consumption of the accelerators in different states—data transfer as sender, data transfer as receiver, kernel execution, and idleness—is described separately and specified by the system specification files.}
For energy optimization, the pipeline’s total energy is assessed by accounting for stage idleness, data transfers, and kernel execution. Accelerator power consumption in states such as data transfer, execution, and idleness is specified in system configuration files.

Once both $dp_{\text{eng}}$ and $dp_{\text{perf}}$ tables are fully populated, we can analyze energy-performance trade-offs by reviewing $dp[|wl|][f][g]$ across various combinations of $f$ and $g$, representing the states after all kernels have been scheduled. Our system, \textsc{DyPe}, automatically explores these configurations based on specified throughput and energy efficiency requirements. For instance, our predefined \textit{balanced} mode allows up to a 30\% reduction in throughput compared to the maximum achievable throughput, aiming to minimize energy consumption.

\setlength{\textfloatsep}{1pt}% Remove \textfloatsep
\begin{algorithm}
\caption{\textsc{DyPe}'s Scheduling Algorithm}\label{alg:scheduling}
\begin{algorithmic}[1]

\footnotesize
\Require $wl,  n_{F}, n_{G}, f_{perf}$ , $f_{comm}$, $f_{eng}$
\State $dp_{perf}[|wl|][n_{F}][n_{G}] \leftarrow {\top}$ 
\State $dp_{perf}[0][0][0] \leftarrow$ empty pipeline with execution time 0
\State $dp_{eng}[|wl|][n_{F}][n_{G}] \leftarrow \top$ 
\State $dp_{eng}[0][0][0] \leftarrow$ empty pipeline with energy 0\\
\textbf{for} {$i \in [1, |wl|]$} \textbf{do} \\
\textbf{for} {$f \in [0, n_{F}]$} \textbf{do} \\
\textbf{for} {$g \in [0, n_{G}]$} \textbf{do} \\
           \textbf{for} {$j \in [1,i]$} \textbf{do} \\
                 \hspace{1.3em} \textbf{for} {$n_f\in [1, f]$} \textbf{do} \\
                    \hspace{0.7em}  \commentgreen{// information about last stage of the previous schedule} \\ 
                     \textcolor{blue}{ \hspace{2.6em} $prev\_stage \leftarrow$ last stage of $dp_{perf}[i$-$j][f$-$n_f][g]$} \\
                    \textcolor{orange}{ \hspace{2.6em} $prev\_stage \leftarrow$ last stage of $dp_{eng}[i$-$j][f$-$n_f][g]$} \\
                    \hspace{2.6em} $src\_dev \leftarrow$ device used in $prev\_stage$ \\
                    \hspace{2.6em} $n_{src\_dev} \leftarrow$ number of devices used in $prev\_stage$ \\
                    \hspace{2.6em} $t_{prev\_stage} \leftarrow$ the execution time of $prev\_stage$ \\
                    \commentgreen{\hspace{2.6em}// data transfer time to the FPGAs} \\
                    \hspace{2.6em} $t^{dst}_{comm_{i\text{-}j}}\leftarrow$ data transfer time to the \textbf{destination} devices, between the kernel $wl[i$-$j]$ and $wl[i$-$j$+$1]$, using $n_{src\_dev}$ units of $src\_dev$ devices and $n_f$ untis of FPGAs, evaluated with $f^t_{comm}$. \\
                    \hspace{2.6em}\commentgreen{// the execution time of the new stage} \\
                    \hspace{2.6em} $t_{new\_stage} \leftarrow f_{perf} (wl[$$i$-$j$:$i$$], $FPGA$, n_f) + t^{dst}_{comm_{i-j}} $ \\
                    \hspace{2.6em}\commentgreen{// data transfer time of the previous stage} \\
                     \hspace{2.6em} $t^{src}_{comm_{i-j}}\leftarrow$ data transfer time to the \textbf{source} devices, between the kernel $wl[i$-$j]$ and $wl[i$-$j$+$1]$, using $n_{src\_dev}$ units of $src\_dev$ devices and $n_f$ untis of FPGAs, evaluated with $f^t_{comm}$. \\
                    \hspace{2.6em}\commentgreen{// update the longest stage} \\
                    \hspace{2.6em} $t_{new\_pipeline} \leftarrow $new longest stage \\
                    \hspace{2.6em}\commentgreen{// update $dp_{perf}$ table} \\
                    \hspace{2.6em} \textcolor{blue}{\textbf{if} {$t_{old\_pipeline} > t_{new\_pipeline}$} \textbf{then} \\
                        \hspace{3.9em} $dp_{perf}[i][f][g] \leftarrow$ new pipeline  \\
                    \hspace{2.6em} \textbf{end if} }\\
                \hspace{2.6em} \commentgreen{// update $dp_{eng}$ table} \\
                \hspace{2.6em} \textcolor{orange}{$new\_pipeline \leftarrow$ add the new stage to $dp_{eng}[i$-$j][f$-$n_f][g]$ \\
                \hspace{2.6em} $e_{new\_pipeline}$ = $f_{eng}(new\_pipeline, t_{new\_pipeline})$ \\
                \hspace{2.6em} \textbf{if} {$dp_{eng}[i][f][g] > e_{new\_pipeline}$} \textbf{then} \\
                    \hspace{3.9em} $dp_{eng}[i][f][g] \leftarrow  new\_pipeline $\\
                \hspace{2.6em} \textbf{end if}}\\
                
                \hspace{1.3em} \textbf{end for}\\
                \hspace{1.3em} \textbf{for} {$n_g\in [0, g]$} \textbf{do}\\
                    \hspace{2.6em} same procedure for GPUs by looking into $dp_{perf}[i$-$j][f][g$-$n_g]$ and $dp_{eng}[i$-$j][f][g$-$n_g]$\\
                \hspace{1.3em} \textbf{end for} \\
\textbf{end for} \\
\textbf{end for} \\
\textbf{end for} \\
\textbf{end for}
\end{algorithmic}
\end{algorithm}
\setlength{\textfloatsep}{1pt}% Remove \textfloatsep

\subsection{\zhen{Data Transfer Costs Estimation}}
\begin{figure}[h] 
\centering
\includegraphics[width=\linewidth]{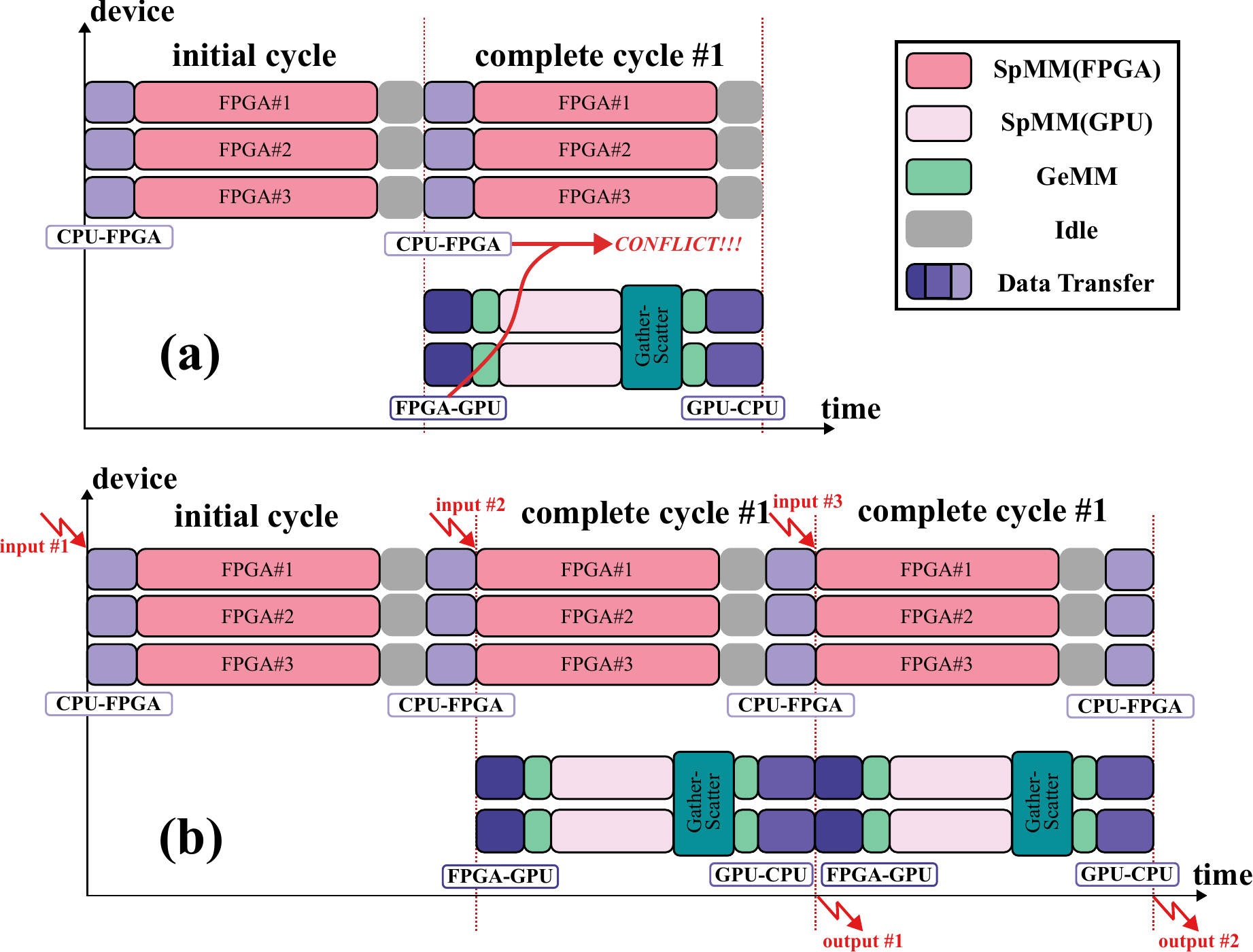}
\caption{Example 2-stages pipeline with and without conflict.}
\label{fig:schedules}
\vspace{-10pt}
\end{figure}

% \textbf{Data Transfer Costs} 
To evaluate data transfer costs effectively, our algorithm uses the $f_{comm}$ function. It is crucial to manage communications between GPU-FPGA and CPU-FPGA pairs to prevent conflicts. Implementing communication processes naively at the onset or conclusion of each computational stage often leads to interference. We maintain a scheduling technique that introduces a delay equivalent to one CPU-FPGA communication cycle at the end of the initial phase, ensuring temporal separation from the subsequent FPGA-GPU data transfer. Handling conflicts is crucial in environments where compute and communication kernels compete for limited resources like HBM and PCIe\textsuperscript{\textregistered} bandwidth, potentially slowing down computations. By avoiding such overlaps, our model accurately predicts latencies for both compute and communication kernels. However, overlaps between CPU-FPGA and GPU-CPU communications are permissible due to their connection to distinct CPUs (see Figure~\ref{fig:schedules}). Figure~\ref{fig:schedules}b illustrates how this scheduling approach aids in deducing a viable schedule.

Moreover, intra-stage data transfers are occasionally necessary when managing input data across multiple devices. For instance, in Graph Neural Network (GNN) workloads, this involves distributing graphs and network data across devices. Our approach employs a data partition strategy, pre-loading static components like graphs and weights onto devices while distributing dynamic elements such as feature matrices and intermediate results during runtime. This reduces the frequency of data transfers and amortizes costs across multiple processing iterations. We incorporate the costs of gather-scatter operations into $f_{perf}$, which assesses stage execution times by considering both the data volume transferred and the available bandwidth for cross-device communication.

